\documentclass[prd,notitlepage,nofootinbib,onecolumn,preprintnumbers]{revtex4-1}
\usepackage{amsmath}
\usepackage{amssymb}
\usepackage{braket}
\usepackage{cancel}
\usepackage{graphicx}
\usepackage{color}
\usepackage[utf8]{inputenc}
\usepackage{hyperref}
\usepackage{multirow}
\usepackage{slashed}
\usepackage[normalem]{ulem}
\usepackage{wasysym}
\usepackage[dvipsnames]{xcolor}

\definecolor{darkgreen}{rgb}{0.0,0.6,0.0}

\newcounter{TODO}

\newcommand{\order}[1]{\mathcal{O}\left(#1\right)}
\newcommand{\refapp}[1]{Appendix~\ref{app:#1}}
\newcommand{\refeq}[1]{Eq.~(\ref{eq:#1})}
\def\eqs#1#2{{Eqs.~(\ref{#1})--(\ref{#2})}}
\newcommand{\refsec}[1]{Section~\ref{sec:#1}}
\newcommand{\reftab}[1]{Table~\ref{tab:#1}}
\newcommand{\reffig}[1]{Figure~\ref{fig:#1}}

\DeclareMathOperator{\sgn}{sgn}

\DeclareMathOperator{\re}{Re}
\DeclareMathOperator{\im}{Im}

\newcommand{\wT}[1]{\widetilde{#1}}

\newcommand{\Afb}{A_\text{FB}}
\newcommand{\thl}{{\theta_\ell}}
\newcommand{\thD}{{\theta_{D}}}

\newcommand{\mDst}{m_{D^*}}

\begin{document}

\preprint{%
    EOS-2021-03,
    TUM-HEP 1305/20,
    P3H-21-021,
    SI-HEP-2021-12
}

\title{\boldmath
Lepton-flavour non-universality of 
\texorpdfstring{$\bar{B}\to D^*\ell \bar\nu$}{B-> D*+ l v}
angular distributions in and beyond the Standard Model
}

\author{Christoph Bobeth}
\email{christoph.bobeth@tum.de}
\affiliation{
  Technische Universit\"at M\"unchen,
  James-Franck-Stra\ss{}e 1,
  D-85748 Garching, Germany
}

\author{Marzia Bordone}
\email{marzia.bordone@to.infn.it}
\affiliation{
  Dipartimento di Fisica,
  Universit\`a di Torino \& INFN, Sezione di Torino,
  I-10125 Torino, Italy
}

\author{Nico Gubernari}
\email{nicogubernari@gmail.com}
\affiliation{
  Technische Universit\"at M\"unchen,
  James-Franck-Stra\ss{}e 1,
  85748 Garching, Germany
}
\affiliation{
  Theoretische Physik 1,
  Naturwissenschaftlich-Technische Fakult\"at,
  Universit\"at Siegen,
  Walter-Flex-Stra{\ss}e 3,
  D-57068 Siegen, Germany
}

\author{Martin Jung}
\email{martin.jung@unito.it}
\affiliation{
  Dipartimento di Fisica,
  Universit\`a di Torino \& INFN, Sezione di Torino,
  I-10125 Torino, Italy
}

\author{Danny van Dyk}
\email{danny.van.dyk@gmail.com}
\affiliation{
  Technische Universit\"at M\"unchen,
  James-Franck-Stra\ss{}e 1,
  D-85748 Garching, Germany
}

\begin{abstract}
We analyze in detail the angular distributions in $\bar{B}\to D^*\ell \bar\nu$
decays, with a focus on lepton-flavour non-universality. We investigate the
minimal number of angular observables that fully describes current and upcoming
datasets, and explore their sensitivity to physics beyond the Standard Model
(BSM) in the most general weak effective theory. We apply our findings to the
current datasets, extract the non-redundant set of angular observables from
the data, and compare to precise SM predictions that include lepton-flavour
universality violating mass effects. Our analysis shows that the current presentation of
the experimental data is not ideal and prohibits the extraction of the full set
of relevant BSM parameters, since the number of independent angular observables
that can be inferred from data is limited to only four. We uncover a $\sim4\sigma$
tension between data and predictions that is hidden in the redundant presentation
of the Belle 2018 data on $\bar{B}\to D^*\ell \bar\nu$ decays. This tension
specifically involves observables that probe $e-\mu$ lepton-flavour universality.
However, we find inconsistencies in these data, which renders results based on
it suspicious.
Nevertheless, we discuss which generic BSM scenarios could explain the tension,
in the case that the inconsistencies do not affect the data materially.
Our findings highlight that $e-\mu$ non-universality in the SM, introduced
by the finite muon mass, is already significant in a subset of angular
observables with respect to the experimental precision.
\end{abstract}

\maketitle

%
%
%
%--------+---------+---------+---------+---------+---------+---------+---------+
\section{Introduction}

Exclusive $\bar{B}\to D^{(*)} \ell \bar\nu$ decays have become precision probes
of the semileptonic parton-level transitions $b\to c \ell \bar\nu$. As such,
they provide excellent means for the determination of the corresponding
Cabibbo-Kobayashi-Maskawa (CKM) matrix element $|V_{cb}|$ of the Standard
Model~(SM). The combination of good experimental and theoretical control renders
them also sensitive probes of beyond-the-SM (BSM) physics that potentially
modifies both the normalization and the angular distribution of these modes.
In the SM, the lepton-flavour universal (LFU) nature of the underlying
$W^\pm$-boson exchange allows for precision predictions of LFU ratios that are
almost free of hadronic uncertainties. Measurements of the three different
lepton modes $\ell=e, \mu, \tau$ then allow  to test SM paradigms such as
CKM unitarity and LFU.
Improved LFU tests are especially important in light of the recent
indications for LFU violation in the so-called \emph{B anomalies}, concerning
$b\to c\tau\bar \nu$ and $b\to s\ell^+\ell^-$ ($\ell=e,\mu$) transitions.
Further motivation for precision analyses of $\bar{B}\to  D^*\ell\bar\nu$
decays is provided by the persisting $V_{cb}$ puzzle, \textit{i.e.} a tension between the
inclusive and exclusive determinations of this CKM element.\\[-0.3cm]

This work is triggered by three recent developments:
\begin{description}
    \item[Availability of experimental data] Starting with the 2015 analysis of
    $\bar B\to D\ell\bar\nu$ decays by Belle~\cite{Glattauer:2015teq},
    experimental collaborations made their data on $b\to c\ell\bar\nu$ transitions
    available in a model-independent way \cite{Glattauer:2015teq,Abdesselam:2017kjf,
    Waheed:2018djm,Aaij:2020hsi,Aaij:2020xjy}, thereby making phenomenological
    analyses possible that vary the  form-factor parametrizations and BSM scenarios.
    In particular, a recent Belle analysis \cite{Waheed:2018djm} presents for the
    first time four single-differential distributions of $\bar B\to D^*\ell\bar\nu$
    decays for both $\ell=e,\mu$ including their full correlation matrices.
    \item[Improved form-factor determinations] There has been significant progress
    in the theoretical determination of hadronic $\bar B\to D^{(*)}$ form factors,
    both from lattice QCD computations \cite{Lattice:2015rga,Na:2015kha,
    Harrison:2017fmw,Bailey:2014tva} and from light-cone sum rules
    \cite{Gubernari:2018wyi}. These determinations allow for precise predictions
    of the complete set of form factors in $\bar B\to D^*\ell\bar\nu$ in the
    whole phase space~\cite{Bordone:2019guc, Bordone:2019vic}. These predictions
    are using the heavy-quark expansion and account for contributions up to and
    including $\order{1/m_c^2}$. They are a prerequisite for a general BSM 
    analysis of these modes.
    \item[Impending progress in experimental and theoretical precision]
    Both the experimental and the theoretical precision are expected to improve
    significantly: the ongoing Belle II and LHCb upgrade experiments are bound
    to deliver $\bar{B}\to D^{(*)}\ell\bar\nu$ results based on multiples of the
    current datasets~\cite{Kou:2018nap, Cerri:2018ypt, Bediaga:2018lhg},
    and updated lattice QCD results for several $\bar{B}\to D^*$ form factors
    beyond zero recoil are upcoming~\cite{Kaneko:2019vkx, Vaquero:2019ary,
    Bhattacharya:2020xyb}, see also the discussions in Refs.~\cite{Lehner:2019wvv,
    Gambino:2020jvv}. This renders the discussion of presently negligible
    effects important for the full phenomenological exploitation of the
    upcoming experimental and theoretical results.
\end{description}
The discussions resulting from the first two items significantly improve our
understanding of these modes, and their sensitivity to the adopted form-factor
parametrization. Recent phenomenological analyses have also shown that the
$V_{cb}$ puzzle is significantly reduced, albeit not yet fully resolved
\cite{Bigi:2016mdz, Bigi:2017njr, Bernlochner:2017xyx, Bigi:2017jbd,
Bernlochner:2017jka, Gambino:2019sif, Bordone:2019guc, Bordone:2019vic,
Bernlochner:2019ldg, Jaiswal:2020wer}. We pose the following questions that
affect existing and future angular analyses of $\bar B\to D^*\ell\bar\nu$ data:
\begin{enumerate}
    \item What is the amount of LFU violation in the SM induced by the muon mass? Is the muon mass still negligible given the achieved experimental and theoretical precision?
    \item What amount of information can be extracted from the available
    single-differential distributions in comparison to a fully-differential
    angular analysis of $\bar B\to D^*\ell\bar\nu$? Is it possible to increase
    the sensitivity to BSM physics with available data by modifying the
    analysis strategy?
    \item What are the limits on BSM physics from existing
    $\bar B\to D^*\ell\bar\nu$ data? Which effective operators could resolve
    a potential tension with the SM and what would be their implications
    on so far unmeasured observables?
\end{enumerate}

In order to answer these questions, we proceed as follows: We begin by describing
the general properties of the $\bar{B}\to D^* \ell\bar\nu$ angular distribution
and the BSM physics reach of the angular observables arising from this distribution
in \refsec{fit-model}. In \refsec{data} we prepare a full angular analysis on
the basis of the Belle data published in Ref.~\cite{Waheed:2018djm}. In doing
so, we identify two obstacles to the full use of these data. In \refsec{fit} we
carry out a fit of the full angular distribution to the Belle data, and discuss
the compatibility with SM predictions. In light of an observed tension, we
further discuss possible BSM interpretations of our results.
We conclude in \refsec{conclusions}.

% 
%
%
%
%--------+---------+---------+---------+---------+---------+---------+---------+
\section{Full angular distribution and its BSM reach}
\label{sec:fit-model}

The four-fold differential distribution of $\bar B\to D^{*}
(\to D\pi) \ell \bar\nu$ decays constitutes a powerful tool for assessing SM
as well as BSM physics. It is given as
\begin{align}\label{eq:d4Gamma}
  \frac{d^4 \Gamma^{(\ell)}}{d q^2\, d\!\cos\thl\, d\!\cos\thD\, d\chi} &
  = \frac{3}{8 \pi} \sum_i J_i^{(\ell)}(q^2) \; 
           f_i(\cos\thl,\, \cos\thD,\, \chi) \,.
\end{align}
Assuming a purely P-wave $D\pi$ final state, this distribution is fully
described by twelve angular observables $J_i^{(\ell)}$ and their respective
angular coefficient functions $f_i$. The dependence of the functions $f_i$
on the three angles $\cos\thl$, $\cos\thD$ and~$\chi$, given in
\refeq{ang-dist} in \refapp{AngularDistribution}, is
lepton-flavour universal and completely determined by conservation of
angular momentum.\\[-0.3cm]

The angular observables $J_i^{(\ell)}$ depend on the momentum
transfer $q^2$, or equivalently the hadronic recoil $w$. Their calculation
involves the lepton-flavour-universal hadronic form factors, as well as the
short-distance coefficients of the low-energy effective theory. The latter
encode short-distance SM effects (which are again lepton-flavour universal)
as well as potential BSM effects (which are in general non-universal). These
dependencies are listed in \reftab{J_i-C_a}. Additional sources of
lepton-flavour non-universality are known kinematic phase-space effects
$\sim m_\ell/\sqrt{q^2}$, which are most pronounced for $\ell = \tau$.
Under the assumption that the short-distance behaviour corresponds to the SM
expectation, the angular observables $J_i^{(\ell)}$ can be used to extract
information on the hadronic form factors. When lifting this assumption in
BSM scenarios, the BSM short-distance coefficients cannot be fully disentangled
from the form factors, making theory input for
the $q^2$-dependence of the form factors and their ratios
indispensable. Below
we discuss the necessary amount of experimental information on the angular
observables $J_i^{(\ell)}$ for a reliable determination of BSM contributions.
Details on the definitions of the angular observables are given in
\refapp{AngularDistribution}.\\

The complete dependence of the angular distribution on BSM contributions
in terms of the BSM couplings has been given for the first time in
Ref.~\cite{Duraisamy:2014sna}, see also Ref.~\cite{Ivanov:2016qtw}, 
with previous partial results throughout the literature \cite{Tanaka:1994ay,
Biancofiore:2013ki,Duraisamy:2013pia,
Fajfer:2012vx,
%Dorsner:2013tla,
Tanaka:2012nw,
%Sakaki:2013bfa}
Korner:1989qb,Hagiwara:1989gza}.
We use the conventions/notation provided in \refapp{AngularDistribution}.
The sensitivity to various BSM couplings and lepton-mass effects have been
studied in detail \cite{Alguero:2020ukk} based on helicity amplitudes.\\[-0.3cm]

Here we would like to address properties that are not mentioned previously,
or that are particularly important for our work.
An important observation in charged-current semileptonic decays is that to
extremely good approximation no CP-conserving scattering phases appear in the
$J_i^{(\ell)}$.\footnote{%
    Such CP-conserving phases are strongly suppressed in $\bar{B}\to D^* \ell \bar\nu$
    and can arise, \emph{e.g.} at the level of dimension eight in the
    low-energy EFT or due to radiative QED corrections.
}
This simplifies their properties under CP conjugation, rendering them simply
even (for $i\in\{1c,1s,2c,2s,3,4,5,6c,6s\}$) or odd (for $i\in\{7,8,9\}$).
As a consequence, the numerators in the combinations 
\begin{align}
    \label{eq:def-S_i-A_i}
    \braket{S_i^{(\ell)}} & 
    \equiv \frac{\braket{J_i^{(\ell)}} + \braket{\bar{J}_i^{(\ell)}}}
                {\Gamma^{(\ell)} + \bar{\Gamma}^{(\ell)}} \,,
&
    \braket{A_i^{(\ell)}} & 
    \equiv \frac{\braket{J_i^{(\ell)}} - \braket{\bar{J}_i^{(\ell)}}}
                {\Gamma^{(\ell)} + \bar{\Gamma}^{(\ell)}} \,,
\end{align}
either vanish or are given by $2\langle J_i^{(\ell)}\rangle$. Here the notation
$\braket{\ldots}$ denotes integration over the full range of the dilepton-invariant
mass as defined in \refeq{def-braket}.

The experimental determination of the fully differential rate is rather
involved. Many analyses therefore present only results for the partially
or fully integrated rate, typically CP-averaged. Doing so simplifies the
experimental analysis, but the sensitivity to some of the angular observables
is lost, which can render the determination of some parameters of interest
impossible. The two recent Belle analyses for instance \cite{Abdesselam:2017kjf,
Waheed:2018djm} provide binned CP-averaged measurements of the four
single-differential distributions
\begin{align}
  \label{eq:dG:dw}
  \frac{d \widehat{\Gamma}^{(\ell)}}{dw} & \equiv 
  \frac{1}{2} \frac{d(\Gamma^{(\ell)} + \bar{\Gamma}^{(\ell)})}{dw} \,,
\\[0.2cm]
\label{eq:dG:dcosthL}
  \frac{1}{\widehat{\Gamma}^{(\ell)}}
  \frac{d \widehat{\Gamma}^{(\ell)}}{d\!\cos\thl} & =
      \frac{1}{2}
    + \braket{\Afb^{(\ell)}} \cos\thl 
    + \frac{1}{4} \left(1 - 3 \braket{\wT{F}^{(\ell)}_L} \right)
      \frac{3 \cos^2 \thl - 1}{2} \,,
\\[0.2cm]
\label{eq:dG:dcosthD}
  \frac{1}{\widehat{\Gamma}^{(\ell)}}
  \frac{d \widehat{\Gamma}^{(\ell)}}{d\!\cos\thD} & = 
     \frac{3}{4} \left(1 - \braket{F_L^{(\ell)}}\right) \sin^2\!\thD
   + \frac{3}{2} \braket{F_L^{(\ell)}} \cos^2\!\thD \,,
\\[0.2cm]
\label{eq:dG:dchi}
  \frac{1}{\widehat{\Gamma}^{(\ell)}}
  \frac{d \widehat{\Gamma}^{(\ell)}}{d\chi} & =  
      \frac{1}{2\pi}
    + \frac{2}{3\pi} \braket{S_3^{(\ell)}} \cos 2\chi 
    + \frac{2}{3\pi} \braket{S_9^{(\ell)}} \sin 2\chi \,,   
\end{align}
where $\widehat{\Gamma}^{(\ell)}$ denotes the CP-averaged decay rate. 
In particular, in Ref.~\cite{Waheed:2018djm} the authors separate the data 
by the light lepton flavours $\ell=e$ and $\ell=\mu$. The three CP-averaged
single-angular distributions depend on only five out of the twelve angular
observables defined in \refeq{d4Gamma}. Out of these five observables,
the CP-averaged $\braket{S_9^{(\ell)}}$ vanishes independently of the BSM
scenario, as discussed above \refeq{def-S_i-A_i}, and is thus not relevant
for our analysis. This leaves the $D^*$-longitudinal polarization fraction
$\braket{F^{(\ell)}_L}$, the lepton forward-backward asymmetry
$\braket{\Afb^{(\ell)}}$, and two further angular observables
$\braket{\wT{F}^{(\ell)}_L}$ and $\braket{S_3^{(\ell)}}$ as independent
observables in the distributions. Within the SM, $\braket{F^{(\ell)}_L}$
and $\braket{\wT{F}^{(\ell)}_L}$ differ by lepton-mass suppressed terms, only.
In a generic BSM scenario, the two observables can further differ due to
contributions from pseudoscalar and tensor operators, see \reftab{J_i-C_a}.
For more details see \refapp{AngularDistribution}.\\[-0.3cm]

The presentation of the data in terms of single-differential distributions
implies that all angular observables are integrated over the full $q^2$ range.
By binning in $q^2$, the data will provide more information about
the BSM couplings through the $q^2$ shape of the angular observables.
In particular, the binned angular observables yield access to more
and independent bilinear combinations of the BSM couplings than the
$q^2$-integrated ones do. Hence, binning the angular observables will constitute
a powerful tool to discriminate between BSM scenarios, as discussed in more
detail below.

The CP asymmetries of the single-differential rates \eqs{eq:dG:dw}{eq:dG:dcosthD}
vanish independently of the BSM scenario. This can be used to validate
the experimental analyses. The CP asymmetry of the $\chi$-dependent rate in
\refeq{dG:dchi} is fully described by the angular observable $A_9^{(\ell)}$.
A measurement of this CP asymmetry could be accomplished with existing datasets
and would provide important information about potential CP-violating BSM effects.

%
%
%--------+---------+---------+---------+---------+---------+---------+---------+
\subsection{Parametrization of BSM Physics}

BSM physics in $\bar{B}\to D^* \ell \bar\nu$ decays
has been investigated, usually based on the assumption of
three light left-handed neutrino flavours below the electroweak scale.
The corresponding most general low-energy effective theory at dimension
six~\cite{Goldberger:1999yh} can be written as~\cite{Jung:2018lfu}
\begin{equation}
    \label{eq:NPlagrangian}
    \mathcal{L}(b\to c\ell\bar\nu)%^\prime)
    = \frac{4 G_F}{\sqrt{2}}V_{cb} \; 
       \sum_i \sum_{\ell'} C_i^{\ell\ell^\prime} \mathcal{O}_i^{\ell\ell^\prime}
    + \text{h.c.} \,.
\end{equation}
Here the operators are constructed out of SM fermion fields
and read
\begin{equation}
    \label{eq:NP_op_basis}
\begin{aligned}
    \mathcal{O}_{V_L}^{\ell\ell^\prime} & 
    =(\bar{c}\gamma^\mu P_L b)(\bar{\ell}\gamma_\mu P_L \nu_{\ell^\prime})\,, 
    \qquad
& 
    \mathcal{O}_{S_L}^{\ell\ell^\prime} &
    = (\bar{c} P_L b)(\bar{\ell} P_L \nu_{\ell^\prime})\,,
    \qquad
&   \mathcal{O}_T^{\ell\ell^\prime} &
    = (\bar{c}\sigma^{\mu\nu} P_L b)(\bar{\ell}\sigma_{\mu\nu} P_L \nu_{\ell^\prime})\,,
\\[0.2cm]
    \mathcal{O}_{V_R}^{\ell\ell^\prime} &
    = (\bar{c}\gamma^\mu P_R b)(\bar{\ell}\gamma_\mu P_L \nu_{\ell^\prime})\,,
&
    \mathcal{O}_{S_R}^{\ell\ell^\prime} &
    = (\bar{c} P_R b)(\bar{\ell} P_L \nu_{\ell^\prime})\,.
\end{aligned}
\end{equation}
They account for lepton-flavour violation (LFV) by $\ell\neq\ell'$.\\

The observables in $\bar{B}\to D^* \ell \bar\nu$ depend only on four
combinations of Wilson coefficients:
\begin{align}
  C_V^{\ell\ell^\prime} & = C_{V_R}^{\ell\ell^\prime} + C_{V_L}^{\ell\ell^\prime} , &
  C_A^{\ell\ell^\prime} & = C_{V_R}^{\ell\ell^\prime} - C_{V_L}^{\ell\ell^\prime} , &
  C_P^{\ell\ell^\prime} & = C_{S_R}^{\ell\ell^\prime} - C_{S_L}^{\ell\ell^\prime} , &
\end{align}%
together with $C_T^{\ell\ell^\prime}$, whereas the combination
$C_S^{\ell\ell^\prime} = C_{S_R}^{\ell\ell^\prime} + C_{S_L}^{\ell\ell^\prime}$
enters only in $\bar{B}\to D \ell \bar\nu$. 
Since the neutrino flavour~$\ell'$ is not detectable, it must be summed over
in every observable. 

\begin{table}
\centering
\renewcommand{\arraystretch}{1.4}
\begin{tabular}{|c|cccc|ccc|cc|c|}
\hline
  Observable & $|C_A|^2$ & $|C_V|^2$ & $|C_P|^2$ & $|C_T|^2$ 
                 & $\re(C_A^{} C_V^*)$ & $\re(C_A^{} C_P^*)$ & $\re(C_A^{} C_T^*)$
                 & $\re(C_V^{} C_P^*)$ & $\re(C_V^{} C_T^*)$ 
                 & $\re(C_P^{} C_T^*)$
\\
\hline\hline
  $J_{1c} = V_1^0$    & $\checkmark$  & --          & $\checkmark$   & $\checkmark$ & -- & $(m)$ & $(m)$ & -- & --    & --\\
  $J_{1s} = V_1^T$    & $\checkmark$  & $\checkmark$& --             & $\checkmark$ & -- & --    & $(m)$ & -- & $(m)$ & --\\
  $J_{2c} = V_2^0$    & $\checkmark$  & --          & --             & $\checkmark$ & -- & --    & --    & -- & --    & --\\
  $J_{2s} = V_2^T$    & $\checkmark$  & $\checkmark$& --             & $\checkmark$ & -- & --    & --    & -- & --    & --\\
  $J_{3}  = V_4^T$    & $\checkmark$  & $\checkmark$& --             & $\checkmark$  & -- & --   & --    & -- & --    & --\\
  $J_{4}  = V_1^{0T}$ & $\checkmark$  & --          & --             & $\checkmark$  & -- & --   & --    & -- & --    & --\\
  $J_{5}  = V_2^{0T}$ & $(m^2)$ & -- & -- & $(m^2)$ & $\checkmark$  & $(m)$ & $(m)$ & -- & $(m)$ & $\checkmark$ \\
  $J_{6c} = V_3^0$    & $(m^2)$ & -- & -- & --      & --            & $(m)$ & $(m)$ & -- & --    & $\checkmark$ \\
  $J_{6s} = V_3^T$    & --      & -- & -- & $(m^2)$ & $\checkmark$  & --    & $(m)$ & -- & $(m)$ & -- \\
\hline
  $d\Gamma/dq^2$  & $\checkmark$ & $\checkmark$ & $\checkmark$ & $\checkmark$ & -- & $(m)$ & $(m)$ & -- & $(m)$ & -- \\
  num($\Afb$)     & $(m^2)$ & -- & --       & $(m^2)$ & $\checkmark$ & $(m)$  & $(m)$ & -- & $(m)$  & $\checkmark$ \\
  num($F_L$)      & $\checkmark$ & -- & $\checkmark$  & $\checkmark$ & -- & $(m)$  & $(m)$ & -- & -- & --\\
  num($F_L$-1/3)      & $\checkmark$ & $\checkmark$ & $\checkmark$  & $\checkmark$ & -- & $(m)$  & $(m)$ & -- & $(m)$ & --\\
  num($\wT F_L$)  & $\checkmark$ & $(m^2)$ & $\checkmark$  & $\checkmark$ & -- & $(m)$ & $(m)$ & -- & $(m)$ & --\\
  num($\wT F_L$-1/3)  & $\checkmark$ & $\checkmark$ & -- & $\checkmark$ & -- & -- & -- & -- & -- & --\\
  num($S_3$)      & $\checkmark$ & $\checkmark$ & --            & $\checkmark$ & -- & -- & -- & -- & -- & --\\
\hline\hline
  Observable & -- & -- & -- & -- 
                 & $\im(C_AC_V^*)$ & $\im(C_AC_P^*)$ & $\im(C_AC_T^*)$
                 & $\im(C_VC_P^*)$ & $\im(C_VC_T^*)$ 
                 & $\im(C_PC_T^*)$
\\
\hline
  $J_{7} = V_3^{0T}$ &  &  &  &  & $(m^2)$      & -- & $(m)$ & $(m)$ & -- & $\checkmark$\\
  $J_{8} = V_4^{0T}$ &  &  &  &  & $\checkmark$ & -- & --    & --    & -- & -- \\
  $J_{9} = V_5^T$    &  &  &  &  & $\checkmark$ & -- & --    & --    & -- & -- \\
\hline
\end{tabular}
\renewcommand{\arraystretch}{1.0}
\caption{The dependence of angular observables on combinations of Wilson 
  coefficients. An entry of $\checkmark$ denotes the presence of this combination.
  An entry of $m^n$ denotes the presence of this term, but with kinematic lepton-mass
  suppression $\propto (m_\ell/\sqrt{q^2})^n$ ($n=1,2$). The
  ``num($\cdot$)'' indicates that only the dependence of the numerator of
  this observable is given. The $V_i^a$ have been introduced in Ref.~\cite{Duraisamy:2014sna}.
}
\label{tab:J_i-C_a}
\end{table}

We determine the minimal number of parameters and their ranges necessary
to parametrize these BSM coefficients for different cases.
Starting from the lepton-flavour conserving case, \refeq{NPlagrangian}
contains five complex parameters $C_i^\ell\equiv C_i^{\ell\ell}$ per
charged-lepton species $\ell$.
In the context of BSM analyses of $\bar{B} \to D^*\ell\bar\nu$, the fact that
matrix elements of the scalar $\bar{c}b$ currents vanish implies that one
can maximally determine four linear combinations out of the five Wilson coefficients.
These four complex coefficients can be parametrized by seven real parameters,
since an overall phase is unobservable, \emph{i.e.} all observables are invariant under a
joint phase rotation $C_i^\ell\to \exp(i\phi^\ell)C_i^\ell$.
For instance, one of the complex coefficients can be chosen real and positive,
which leaves four real and three imaginary parts or four absolute values and three relative phases
as free parameters.
The Lagrangian \refeq{NPlagrangian} is conveniently normalized to $G_F\, V_{cb}$
to ensure that in the SM $C_{V_L} = 1$ at tree-level.
In general, these factors cannot
be separated from the BSM Wilson coefficients since only their products
enter observables.
Hence, they do not count as additional parameters.
The set of seven real parameters is therefore the maximal information
we can hope to extract from $\bar B\to D^*\ell\bar\nu$ decays for a given $\ell$ without LFV.\\[-0.3cm]

All CP-averaged observables depend on these seven parameters through
the combinations
\begin{equation}
    \label{eq:LFC-comb-1}
    \begin{aligned}
        |C_i^\ell|^2
            & =\re^2(C_i^\ell) + \im^2(C_i^\ell)\,,\\
        \re(C_i^\ell C_j^{\ell*})
            & = \re(C_i^\ell)\re(C_j^\ell) - \im(C_i^\ell)\im(C_j^\ell)\,.
    \end{aligned}
\end{equation}
These combinations, however, are invariant under the discrete symmetry transformation
$\im(C_i^\ell) \to -\im(C_i^\ell) \;\forall\; i$.
The combinations
\begin{equation}
    \label{eq:LFC-comb-2}
    \im(C_i^\ell C_j^{\ell*})
             = \im(C_i^\ell)\re(C_j^\ell) - \re(C_i^\ell)\im(C_j^\ell)\,,
\end{equation}
can therefore still be determined from CP-averaged observables, albeit only
up to an overall sign.
One is free to choose one of these signs freely in the fit, since the second
solution can always be obtained by inverting the signs of the imaginary
parts.\\[-0.3cm]

In the limit of a massless lepton, the two classes of Wilson coefficients
$C_{A,V}^\ell$ and $C_{P,T}^\ell$ decouple in the observables, since their
interference is $m_\ell$ suppressed, as shown in \reftab{J_i-C_a}.
As we will see below, this applies only to electrons, since in precision
analyses the muon mass cannot be neglected anymore. This implies a separate
symmetry for each class, $C_{V,A}^\ell\to \exp(i\phi^\ell)C_{V,A}^\ell$
and $C_{P,T}^\ell\to \exp(i\varphi^{\ell})C_{P,T}^\ell$.
Therefore another phase cannot be determined from any $\bar{B}\to D^*\ell^-\bar\nu$
observable in this limit.
In fact, it can be eliminated altogether from the parametrization, leaving
maximally six parameters to be determined from $\bar B\to D^*\ell\bar \nu$
for massless charged leptons. In this case also the discrete
symmetry for the imaginary parts holds separately for each class, allowing
to choose another sign freely. Hence, the most general parametrization of
CP-averaged $\bar B\to D^* e\bar\nu$ data within the weak effective theory
and when neglecting LFV requires only six parameters, four of which can be
chosen positive. Taking into account lepton-mass effects
requires a seventh parameter, and only two of these parameters can be chosen positive.\\[-0.3cm]

Note that in the counting above we have assumed the couplings for the different
lepton flavours to be completely independent, allowing in particular for
independent phase rotations.
Such an assumption does not hold in all BSM scenarios; in particular
it does not hold in the
Standard Model Effective Field Theory (SMEFT) at mass dimension six.
In the matching of \refeq{NPlagrangian} to the SMEFT, the coefficient $C_{V_R}^{\ell\ell'}$
is lepton-flavour universal, a property inherited from the SM gauge group \cite{Cata:2015lta, Cirigliano:2009wk}.
This universality couples the different sectors and
consequently the phase rotations cannot be performed independently anymore.
This gives rise to 
an additional measurable phase in this scenario, and therefore necessitates a new corresponding
parameter. 
For instance, for the common and convenient choice of a real and positive
$C_{V_L}^\ell$, the coefficients $C_{V_R}^{\ell}$
cannot be trivially identified with each other. Instead they fulfill
\begin{align}\label{eq::CVRuni}
    C_{V_R}^e & = \exp(i\phi_L) \, C_{V_R}^\mu\,,
&
    \phi_L & = \phi_{V_L}^e-\phi_{V_L}^\mu\,,
\end{align}
and similarly for $\ell=\tau$.
The relative phase between the two Wilson coefficients $C_{V_L}^e$ and
$C_{V_L}^\mu$ appears explicitly, while it can be absorbed everywhere else.
This implies that although two real parameters are removed (one of the complex
$C_{V_R}^\ell$ coefficients), one is added (the relative phase),
and hence the overall number of parameters is reduced only by one.

Generalizing the above observations in the presence of lepton-flavour-violating
interactions, $\ell\neq \ell'$, is straight-forward insofar as the
contributions with different neutrino flavours do not interfere. Hence all
expressions in \eqs{eq:LFC-comb-1}{eq:LFC-comb-2} remain valid with the
generalizations
\begin{align}
  \label{eq:LFV-generalization}
  \left|C_i^\ell\right|^2 &
  \to \sum_{\ell'} \left|C_i^{\ell\ell'}\right|^2 ,
& 
  \re\left(C_i^\ell C_j^{\ell*}\right) &
  \to \sum_{\ell'} \re\left(C_i^{\ell\ell'} C_j^{\ell\ell'*}\right) ,
& 
  \im\left(C_i^\ell C_j^{\ell*}\right) &
  \to \sum_{\ell'} \im\left(C_i^{\ell\ell'} C_j^{\ell\ell'*}\right) .
\end{align}
The symmetry considerations hold for each neutrino flavour separately.
Naively the number of parameters simply triples compared to
the lepton-flavour-conserving case above. The situation is nevertheless
significantly different from the lepton-flavour conserving case, for which the number
of parameters is smaller than the number of combinations of Wilson coefficients
appearing in the description of the decay.
This implies (non-linear) relations
between these combinations in the lepton-flavour conserving case, for instance,
\begin{equation}
    \label{eq:LFV-relation}
    \im^2(C_i^\ell C_j^{\ell*})
        = |C_i^\ell|^2 |C_j^{\ell}|^2-\re^2(C_i^\ell C_j^{\ell*})\,.
\end{equation}
With the generalizations in \refeq{LFV-generalization}, the number of
BSM parameters is larger than the number of combinations of Wilson
coefficients. Hence, the latter determine the maximal number of parameters
(parameter combinations) that can be extracted.
This implies that relations such as \refeq{LFV-relation} do not hold anymore in the presence of lepton-flavour violation and can be used instead
to test for LFV in charged-current decays without the need to identify
the neutrino flavour experimentally.\\[-0.3cm]

In the presence of light right-handed neutrinos, similar considerations
as for the LFV case apply, since also here more BSM parameters are
introduced and the corresponding contributions do not interfere.
The generalization to light right-handed neutrinos is therefore analogous to \refeq{LFV-generalization}
and similar comments apply for the determination of the corresponding
parameters.

%
%
%--------+---------+---------+---------+---------+---------+---------+---------+
\subsection{\boldmath BSM reach in 
\texorpdfstring{$\bar{B}\to D^*\ell \bar\nu$}{B -> D*+ l v}}

We now turn to the determination of the discussed parameters from the
differential distributions. Each fully $q^2$-integrated angular observable provides
one linear combination of the combinations of Wilson coefficients only, as indicated
in \reftab{J_i-C_a}. The measurement of their $q^2$ dependence
allows further to separate different BSM contributions to the same angular observable,
if their $q^2$ dependence \cite{Alguero:2020ukk} is different. For instance,
the $q^2$-differential
rate allows to determine all four absolute values of the BSM parameters.
The question is what amount of
experimental information is necessary to determine the maximal amount
of parameters in a given scenario. \reftab{BSMinfo} shows the situation
in a few scenarios for different sets of experimental measurements.\\[-0.3cm]

\begin{table}[t]
    \centering
    \renewcommand{\arraystretch}{1.4}
\begin{tabular}{l c c c c l}
    \hline\hline
                                 & \multicolumn{2}{c}{no LFV} & \multicolumn{2}{c}{LFV} & \\
    Measurement                 & $m_\ell\to0 \quad$ & $m_\ell>0 \quad$ & $m_\ell\to0 \quad$ & $m_\ell>0 \quad$  & Comments \\
    \hline
        4-fold differential, S+A            & 6                  & 6+1              & 8                  & 8+5              & Maximum achievable\\
        4-fold differential, S only         & 6                  & 6+1              & 6                  & 6+3              & $\sgn[\im(C_i^\ell C_j^{\ell *})]$ not resolved\\
        $4\times1$-fold differential, S+A   & 6                  & 6+1              & 6                  & 6+3              & 2-fold ambiguity in $m_\ell\to0$\\
        $4\times1$-fold differential, S only& 5                  & 5+2              & 5                  & 5+3              & Insufficient for $m_\ell\to 0$,\\
                                            & & & & & $\sgn[\im(C_i^\ell C_j^{\ell*})]$ not resolved\\
    \hline\hline
    \end{tabular}
    \renewcommand{\arraystretch}{1.0}
    \caption{Amount of BSM physics information that can be extracted in
    different scenarios, see also text. Here S and A denote the measurement
    of the CP average and the CP asymmetry of the respective differential rate.
    The first and second number corresponds to the number of parameters that can be
    extracted without and with mass suppression, respectively.}
    \label{tab:BSMinfo}
\end{table}

A few general comments are in order:
\begin{itemize}
    \item It is necessary to consider the CP-conjugated modes separately
    if the sign ambiguity for the imaginary parts is to be resolved.
    Since the lepton charge tags the $B$ meson flavour, this is not difficult
    to achieve experimentally.
    \item The interference between the two classes of BSM coefficients
    $C_{A,V}^\ell$ and $C_{P,T}^\ell$ is always lepton-mass suppressed, see
    \reftab{J_i-C_a}. Hence its determination requires high statistical power,
    as expected from the upcoming datasets at Belle 2 and the LHC experiments.
    \item While for $\ell=\mu$ there is some sensitivity to additional combinations
    of Wilson coefficients, these combinations are still strongly suppressed.
    The corresponding parameters will therefore be determined
    comparatively poorly. Generally the best chance to determine them is to
    consider rather low values of $q^2$, given the suppression by
    powers of $m_\ell/\sqrt{q^2}$, both for the angular
    observables and the $q^2$-differential rate. Probing different bins in~$q^2$
    can also improve the sensitivity to other BSM coefficients. Tensor
    interactions for instance can be probed particularly well at low $q^2$
    in $d\widehat\Gamma_T/dq^2\sim 3S_{1s}-S_{2s}$, since the SM
    contributions vanish for $q^2\to 0$, while the tensor contributions
    remain finite \cite{Jung:2018lfu}, see also Ref.~\cite{Bhattacharya:2015ida}.
\end{itemize}

Considering some of the scenarios in more detail, we make the following
observations:
\begin{itemize}
    \item It is impossible to determine the full set of physical BSM
    parameters for $m_\ell\to0$ (\emph{e.g.} $\ell=e$) from the CP-averaged
    single-differential rates alone, even disregarding ambiguities
    in the signs of imaginary parts. The reason is that in this case only
    $\braket{\Afb^{(\ell)}}$ is sensitive  to the relative
    phases between the coefficients.
    Since there are two observable relative phases (one between $C_A^\ell$
    and $C_V^\ell$, one between $C_P^\ell$ and $C_T^\ell$), they cannot
    both be determined from this single observable. 
    \item Assuming the flavour-conserving case, the extraction of all seven parameters
    is possible for finite $m_\ell$ from the CP-averaged single-differential rates,
    modulo discrete ambiguities. However, one relative phase can only be obtained from
    lepton-mass-suppressed contributions, even though in more sophisticated
    measurements it would be accessible without lepton-mass suppression.
    \item Beyond the lepton-flavour-conserving case, it becomes
    clearer how much more information is contained in a fully $q^2$-differential
    measurement. Strictly speaking, such a measurement is not necessary when
    assuming lepton-flavour conservation. However, also in this case there are
    additional crosschecks possible and additional form-factor information
    can be extracted together with the BSM parameters.
\end{itemize}
These observations apply fully to the recent Belle measurements \cite{Waheed:2018djm}.\\[-0.3cm]

Considering the determination of the full  BSM information in the
lepton-flavour-conserving case as an important intermediate goal, there are
several ways this could be achieved with existing data, extending the experimental
analyses only slightly:
\begin{enumerate}
    \item Measuring $\Afb^{(\ell)}$ in at least two $q^2$ bins. This disentangles
    $S_{6s}^{(\ell)}$ from $S_{6c}^{(\ell)}$ entering this observable. Given that the
    $\cos\theta_\ell$-differential distribution \eqref{eq:dG:dcosthL}
    has been measured in 10 bins in
    Refs.~\cite{Waheed:2018djm, Abdesselam:2017kjf}, but contains only two
    angular observables, this seems feasible by reducing the number of
    $\cos\theta_\ell$ bins and providing the observables in two or
    three $q^2$ bins instead.
    This would give access to all BSM parameters, leaving only two signs of
    imaginary parts undetermined.
    \item Measuring $d\Gamma/d\chi$ separately for the two lepton charges.
    This would give access to $A_9^{(\ell)}$, and thereby to
    $\im(C_A^\ell C_V^{\ell *})$. This in turn would determine also $\re(C_A^\ell
    C_V^{\ell *})$ up to a sign, and thereby allow to access $\re(C_P^\ell
    C_T^{\ell *})$ from $\braket{\Afb^{(\ell)}}$ up to a two-fold ambiguity.
    Each of the solutions would still have a two-fold sign ambiguity for the
    corresponding imaginary part. Together with the first option, this measurement
    would resolve the sign ambiguity in $\im(C_A^\ell C_V^{\ell *})$, leaving
    only the one in $\im(C_P^\ell C_T^{\ell *})$ (should this parameter combination
    be found to be different from zero).
    \item Assessing $S_5^{(\ell)}$, $A_7^{(\ell)}$ and/or $A_8^{(\ell)}$.
    The measurement of each of these requires a different binning
    scheme, since these observables do not enter the single-differential
    rates. The latter two further require tagging by the lepton charge. 
    Of particular interest is $A_7^{(\ell)}$, since contributions linear in BSM
    parameters are additionally lepton-mass suppressed, rendering the quadratic BSM
    contributions  potentially dominant.
    A similar statement holds for~$S_{6c}^{(\ell)}$.
\end{enumerate}

%
%
%
%--------+---------+---------+---------+---------+---------+---------+---------+
\section{Available Experimental Data}
\label{sec:data}

Semileptonic $\bar{B}\to D^{(*)}\ell\bar\nu$ decays have been of key interest for many years,
see Ref.~\cite{Amhis:2019ckw} for a list of analyses over the last
$\sim 25$~years. However, until recently, almost all experimental analyses have been
tied to a specific form-factor parametrization, specifically the so-called
CLN~parametrization \cite{Caprini:1997mu}. This parametrization involves
assumptions that are not adequate anymore for precision analyses. Applying
instead the underlying formalism of a heavy-quark expansion more consistently
\cite{Bernlochner:2017jka} and extending it to include $1/m_c^2$ contributions
\cite{Bordone:2019guc, Bordone:2019vic}, allows for a consistent description
of the available experimental data and form factor results. However, since
experimental analyses presented in most cases \emph{only} parametrization-specific
results, a model-independent reanalysis under different theory assumptions
of the underlying experimental data is impossible.\footnote{The total branching
ratio results have been found to be approximately parametrization-independent,
see, \emph{e.g.} Refs.~\cite{Bigi:2017jbd, Waheed:2018djm}, but might still
suffer from underestimated uncertainties to some extent.} Unfortunately, this
problem persists in the most recent BaBar analysis~\cite{Dey:2019bgc}, which
includes a second form factor parametrization, but still does not allow for an
independent analysis of the data. Furthermore, in many cases electron and muon
data have been averaged without presenting separate results, rendering them of
limited use for the analysis of LFU. A notable exception among these past studies
is the 2010 untagged Belle analysis \cite{Dungel:2010uk}, which presented
lepton-specific differential rates separately for longitudinal and transverse
$D^*$ polarizations, but lacked the necessary correlations.\\[-0.3cm]

More recently, the Belle analysis of $\bar{B}\to D\ell\bar\nu$~\cite{Glattauer:2015teq}
presented for the first time lepton-specific differential rates including
their full correlations, which made possible precision studies with arbitrary
form-factor parametrizations for the first time, initiating an intense ongoing
discussion regarding the best way to analyze these and similar data.
Similar comments apply to the preliminary $\bar{B}\to D^*\ell\bar\nu$ data with hadronic tag in
Ref.~\cite{Abdesselam:2017kjf}, which were however again lepton-flavour averaged
and are presently reanalyzed, and the 2018 untagged analysis \cite{Waheed:2018djm},
superseding the results of Ref.~\cite{Dungel:2010uk}, which we discuss in detail in the following. 

%
%
%--------+---------+---------+---------+---------+---------+---------+---------+
\subsection{Belle's 2018 untagged analysis}
\label{sec:BelleData}

The dataset for the angular distribution provided by Belle~\cite{Waheed:2018djm}
is the first analysis that separates the electron mode from the muon mode in both
the bin contents and the statistical covariance matrix, and also the systematic
covariance matrix can be reconstructed for both lepton species
separately.\footnote{%
    Note that the arXiv version v3 of Ref.~\cite{Waheed:2018djm}
    contains erroneous statistical and systematic correlation matrices. The journal
    version of Ref.~\cite{Waheed:2018djm} contains the correct statistical correlation matrix, 
    but still an erroneous systematic one: both off-diagonal $20\times 20$ blocks 
    of the $40\times 40$ matrix should be transposed.
}
Unfortunately the correlations between the electron and muon modes are not given
explicitly. Yet Belle has used these data for a high-precision LFU test
that compares
the branching fractions to electrons and muons integrated over the entire
phase space. They found the ratio to be in agreement with lepton flavour universality,
$R_{e/\mu} = 1.01 \pm 0.01\mbox{(stat.)} \pm 0.03\mbox{(sys.)}$.
In our study we aim to extend the study of LFU to the angular observables using the same Belle data. For
this purpose we need to construct a combined correlation matrix for the full dataset, including correlations between electrons and muons.\\[-0.3cm]

Before going into these details, however, we comment on an issue present in the
statistical correlation matrix. Belle provides the number of (background-subtracted)
events before unfolding in bins of the four aforementioned single-differential
distributions. These are the distribution in
\begin{align}
    w & = \frac{m_B^2 + \mDst^2 - q^2 }{2 m_B \mDst} ,
\end{align}
and the three angular distributions Eqs.~\eqref{eq:dG:dcosthL}--\eqref{eq:dG:dchi}.
There are 10 equidistant bins for each distribution, resulting in 40 bins per
lepton flavour (LF). The events in the 10 bins for each of the four distributions
sum up to the same number for each lepton flavour:
\begin{equation}
   \label{eq:obs-norm}
    \sum_{i= 1}^{10} N^\text{obs}_{i,\ell}
    = \sum_{i=11}^{20} N^\text{obs}_{i,\ell}
    = \sum_{i=21}^{30} N^\text{obs}_{i,\ell}
    = \sum_{i=31}^{40} N^\text{obs}_{i,\ell}
    = \left\{ \begin{array}{cc} 90743.4 & \ell = e \\[0.2cm]
                                89087.0 & \ell = \mu 
              \end{array} \right. ,
\end{equation}
\emph{i.e.} the same signal candidates have been histogrammed in four different ways
in the four single-differential distributions.
These relations imply that for both electrons and muons only 37 of the measured
bins are independent, since the content of 3 bins can be calculated as the total
yield minus the yields of the other 9 bins of the corresponding distributions.
This in turn implies that the corresponding statistical correlation matrices have
to be singular; each of the $40\times 40$ matrices should exhibit three vanishing
eigenvalues. This is, however, not the case: the determinant of both matrices is
rather large and all eigenvalues of both statistical correlation matrices are
$\mathcal O(1)$. It remains unclear why the statistical correlation matrices do
not reflect the linear dependence of the 3 bins, which should by construction
be a result of the description of 10 bins per single-differential distribution
used by the Belle collaboration. Note that the issue of the linearly dependent
bins affects the determination of $V_{cb}$ from these data:\footnote{%
    Depending on
    the source of this problem, it might not affect the determination of $V_{cb}$
    directly from the electron and muon event samples in Ref.~\cite{Waheed:2018djm}.
}
if the sum over each set of 10 bins is identical, no information is added to the
determination of the total rate by having the four binnings. However, if the
correlations are such that these sums become effectively independent, the total
rate is more precisely determined by considering all four binnings than by
considering only a single one, leading to an underestimation of the uncertainty
of the total rate (and hence $V_{cb}$). 
The effect is not large with the given data, but it is non-vanishing: the determination 
of the total rate is a couple of per mil better than from each individual distribution.
It is important to note that this small numerical impact is \emph{not} an
indication that a correct extraction of the statistical correlation matrices
will lead to small corrections in the analysis. Since there is an unknown problem
in the extraction of the statistical correlation matrices, there is no way of
knowing what the effect of its resolution will be. Given this numerical smallness
within our analysis, however, we will work below with $40 \times 40$ matrices.
In LF-specific fits with a $37\times 37$ matrix the result varies very slightly,
depending on the choice of the three discarded bins, and any specific choice
would be arbitrary. We have checked that our numerical results below remain 
essentially unaffected. The issue with the statistical correlation matrices
must be kept in mind when interpreting \emph{any} results obtained from the
data from Ref.~\cite{Waheed:2018djm}.\\[-0.3cm]

In the remainder of this section, we describe the construction of a combined
electron-muon $80\times 80$ covariance matrix based on Ref.~\cite{Waheed:2018djm},
with only one mild additional assumption. According to Ref.~\cite{Waheed:2018djm},
the only source of systematic uncertainties that is different for $\ell=e$ and
$\ell=\mu$ is the procedure of lepton identification (Lepton ID). Given the
statistical  independence of electron and muon samples, this implies the
following form for the total covariance matrix:
\begin{align}
    \text{Cov}^\text{total}_{80\times 80} &
    = \text{Cov}^\text{stat}_{80\times 80}+\text{Cov}^\text{sys}_{80\times 80}
\\[0.2cm] &
    = \begin{pmatrix}
     \text{Cov}^{\text{stat},e}_{40\times 40} & 0_{40\times 40}
     \\[0.2cm]
     0_{40\times 40} & \text{Cov}^{\text{stat},\mu}_{40\times 40}
    \end{pmatrix}
    + \begin{pmatrix}
     \text{Cov}^{\text{sys,uni}}_{40\times 40} & \text{Cov}^{\text{sys,uni}}_{40\times 40}
     \\[0.2cm]
     \text{Cov}^{\text{sys,uni}}_{40\times 40} & \text{Cov}^{\text{sys,uni}}_{40\times 40}
    \end{pmatrix}
    + \text{Cov}^\text{sys,lep-ID}_{80\times 80}\,.
\end{align}
The lepton-ID systematic uncertainties are provided individually for both
lepton flavours, but also for the ``LF-combined'' which enter the
systematic correlation matrix given explicitly in the article. We therefore
have
\begin{align}
    \text{Cov}^\text{sys,uni}_{40\times 40} & 
    = \text{Cov}^\text{sys,LF-comb}_{40\times 40}
    - \text{Cov}^\text{sys,lep-ID-comb}_{40\times 40} \,.
\end{align}
Together with the information that the Lepton-ID systematic uncertainties
are 100\% positively correlated throughout all bins \cite{Waheed-PhD},
$\text{Cov}^{\text{sys,lep-ID-comb}}_{ij} = \sigma_i^{\text{lep-ID-comb}}
\sigma_j^{\text{lep-ID-comb}}$, where $\sigma_i$ are systematic uncertainties
of the $i$th bin taken from tables XI--XIV \cite{Waheed:2018djm}, the ``LF
combination'' can thus be undone for the systematic correlations. We compute
the LF-specific systematic covariances
($\text{Cov}^{\text{sys},\ell}$) from the ``LF-combined'' ones
($\text{Cov}^\text{sys,LF-comb}$) of \cite{Waheed:2018djm} consequently as
\begin{align}
   \text{Cov}^{\text{sys},\ell}_{ij} & 
   = \text{Cov}^\text{sys,LF-comb}_{ij} 
   - \sigma_i\sigma_j |^{\text{lep-ID-comb}}
   + \sigma_i\sigma_j |^{\text{lep-ID,}\ell} ,
&
   i,j & = 1,\ldots, 40\,.
\end{align}
LF-specific analyses can be performed with these LF-specific $40 \times 40$
statistical and systematic correlation matrices at hand.
The \emph{only} assumption we make for the construction of the full
$80\times 80$ covariance matrix is that the lepton-ID uncertainties
for electrons and muons are uncorrelated:
\begin{align}
    \text{Cov}^\text{sys,lep-ID}_{80\times 80} 
    = \begin{pmatrix}
    \text{Cov}^{\text{sys,lep-ID},e}_{40\times 40} & 0_{40\times 40}
    \\[0.1cm]
    0_{40\times 40} & \text{Cov}^{\text{sys,lep-ID},\mu}_{40\times 40}
    \end{pmatrix}\,.
\end{align}
This is plausible (as confirmed by Belle collaboration members \cite{PrivateCommunications}), given
they concern different detector parts, but not fully guaranteed.
We consider this assumption to be at a comparable level to the assertion
in Ref.~\cite{Waheed:2018djm} that the lepton ID constitutes the only
non-universal contribution to the systematic uncertainty. Note that this
is an approximation that might not hold well enough to analyze LFU.
In that case the systematic uncertainty given in \cite{Waheed:2018djm}
for the LFU ratio $R_{e/\mu}$ would be underestimated, as would be our $e-\mu$
covariance. However, we perform below an extremely conservative check
that our observation of a tension with the SM does not depend on this
assumption.

%
%
%
%--------+---------+---------+---------+---------+---------+---------+---------+
\section{\boldmath Fits to 
\texorpdfstring{$\bar{B}\to D^* (e, \mu)\bar\nu$}{B-> D*+ (e, mu) v}
Data and Discussion}
\label{sec:fit}

We analyze the data from the Belle analysis \cite{Waheed:2018djm} in detail,
based on the general analysis in \refsec{fit-model} and the covariance
matrix derived in \refsec{data}. 

%
%
%--------+---------+---------+---------+---------+---------+---------+---------+
\subsection{Angular analysis and comparison with the SM}

In the first step our fit is completely model-independent: we use the observation
made in \refsec{fit-model} that the three single-differential CP-averaged
angular distributions can be fully described by only four angular observables 
\begin{align}
    \label{eq:obs-set-1}
    \braket{\Afb^{(\ell)}}, \quad 
    \braket{F_L^{(\ell)}}, \quad 
    \braket{\wT F_L^{(\ell)}}, \quad
    \braket{S_3^{(\ell)}} ,
\end{align}
retaining all information. Further, we parametrize the 10 bins of
the $w$-distribution again in full generality as the
total decay rate and nine independent bins of the normalized $w$-differential rate:
\begin{align}
   \label{eq:obs-set-2}
   \widehat{\Gamma}^{(\ell)}, &
&
   x_i^{(\ell)} & \equiv \frac{1}{\widehat{\Gamma}^{(\ell)}}
   \int_{w_{i-1}}^{w_i} \!\! dw \, \frac{d \widehat{\Gamma}^{(\ell)}(w)}{dw} ,
&
   w_i & = 1 + i \frac{(w_\text{max}-1)}{10}
&  (i & = 2, \ldots 10)\,.
\end{align}
Here $w_\text{max} = 1.5$ to comply with the choice in \cite{Waheed:2018djm},
which excludes a tiny part of the low-$q^2$ phase space.
From this parametrization we calculate the bin contents $N_{i,\ell}^\text{obs}$
by integrating over the relevant angle intervals where necessary, and folding
these predictions with the corresponding response matrices and efficiencies provided
by the Belle collaboration for each lepton flavour separately, as described in
\cite{Waheed:2018djm}. We thus arrive at a description of the 40 bins per lepton
flavour given in \cite{Waheed:2018djm} in terms of only $10 + 4 = 14$ observables
in \eqs{eq:obs-set-1}{eq:obs-set-2}.
We emphasize that our fit parameters appear up to the common normalization factor
linearly, assuring a unique minimum and no distortion of their distributions
from a multivariate gaussian shape.\\[-0.3cm]

The conversion of number of events to decay rate involves the following
numerical input:
\begin{equation}
\begin{aligned}
    N_{B\bar{B}} & = (772 \pm 11) \cdot 10^6 , \qquad\qquad
&
    \mathcal{B}(D^{*+} \to D^0 \pi^+) & = (67.7 \pm 0.5)\, \% ,
\\
    f_{00} & = 0.486 \pm 0.006 ,
&
    \mathcal{B}(D^0 \to K^- \pi^+)     & = (3.950 \pm 0.031)\, \% ,
\\
    \tau_{B^0} & = (1.519 \pm 0.004) \cdot 10^{-12} \,\text{s} ,
\end{aligned}
\end{equation}
with $N_{B\bar{B}}$ from \cite{Natkaniec:2006rv}, $f_{00}$ and from the
$B^0$~lifetime from \cite{Amhis:2019ckw} (see also the discussion on
$f_{00}$ in \cite{Jung:2015yma}), and the latest values of the branching
fractions from \cite{Zyla:2020zbs}. Note that the value for $\mathcal{B}(D^0 \to K^- \pi^+)$
was updated w.r.t the value used in Ref.~\cite{Waheed:2018djm, Ferlewicz:2020lxm},
which slightly
impacts the determination of $V_{cb}$. The corresponding uncertainties
cancel in all ratios and hence affect only the total decay rate, for
which they are included in the systematic uncertainties provided by the
Belle collaboration \cite{Waheed:2018djm}.\\[-0.3cm]

We further introduce the averages and differences of LF-specific observables
\begin{align}
    \label{eq:def-LF-ave-obs}
    \Sigma X & \equiv \frac{X^{(\mu)} + X^{(e)}}{2} \,,
&
    \Delta X & \equiv X^{(\mu)} - X^{(e)} \,,
\end{align}
for later convenience in the study of LFU violation where $X^{(\ell)}$ stands
for any of the considered observables.\\[-0.3cm]

\begin{table}
\centering
\renewcommand{\arraystretch}{1.4}
\begin{tabular}{|c|cc|cc|}
\hline
    & \multicolumn{2}{c|}{SM} & \multicolumn{2}{c|}{Fit Belle data}
\\
\hline
    observable & $\ell = e$ & $\ell = \mu$
               & $\ell = e$ & $\ell = \mu$
\\
\hline\hline
    $10^{14} \cdot \widehat{\Gamma}^{(\ell)}$ [GeV]  &
    $(1.469 \pm 0.079)\cdot 10^{3}\cdot|V_{cb}|^2$  &
    $(1.465 \pm 0.078)\cdot 10^{3}\cdot|V_{cb}|^2$  &
    $ 2.1840 \pm 0.0682$ &
    $ 2.1656 \pm 0.0719$
\\
    $x_2^{(\ell)}$ &
    $0.0920 \pm 0.0034$ & $0.0923 \pm 0.0034$ &
    $0.0956 \pm 0.0022$ & $0.0915 \pm 0.0023$
\\
    $x_3^{(\ell)}$ &
    $0.1085 \pm 0.0030$ & $0.1088 \pm 0.0030$ & 
    $0.1146 \pm 0.0024$ & $0.1120 \pm 0.0025$
\\
    $x_4^{(\ell)}$ &
    $0.1161 \pm 0.0020$ & $0.1164 \pm 0.0020$ & 
    $0.1184 \pm 0.0027$ & $0.1208 \pm 0.0029$
\\
    $x_5^{(\ell)}$ & 
    $0.1180 \pm 0.0010$ & $0.1184 \pm 0.0010$ & 
    $0.1198 \pm 0.0033$ & $0.1241 \pm 0.0033$
\\
    $x_6^{(\ell)}$ &
    $0.1160 \pm 0.0006$ & $0.1163 \pm 0.0006$ & 
    $0.1169 \pm 0.0037$ & $0.1134 \pm 0.0037$
\\
    $x_7^{(\ell)}$ & 
    $0.1108 \pm 0.0015$ & $0.1110 \pm 0.0015$ & 
    $0.1079 \pm 0.0040$ & $0.1145 \pm 0.0040$
\\
    $x_8^{(\ell)}$ &
    $0.1031 \pm 0.0025$ & $0.1033 \pm 0.0025$ & 
    $0.1045 \pm 0.0039$ & $0.0942 \pm 0.0040$
\\
    $x_9^{(\ell)}$ &
    $0.0934 \pm 0.0033$ & $0.0936 \pm 0.0033$ & 
    $0.0923 \pm 0.0036$ & $0.0923 \pm 0.0037$
\\
    $x_{10}^{(\ell)}$ &
    $0.0820 \pm 0.0040$ & $0.0818 \pm 0.0040$ & 
    $0.0758 \pm 0.0033$ & $0.0777 \pm 0.0034$
\\
    $\braket{F_L^{(\ell)}}$ &
    $0.541  \pm 0.011$  & $0.542  \pm 0.012$  & 
    $0.5336 \pm 0.0045$ & $0.5271 \pm 0.0046$
\\
    $\braket{\Afb^{(\ell)}}$ & 
    $0.204  \pm 0.012$  & $0.198  \pm 0.012$  & 
    $0.1951 \pm 0.0069$ & $0.2300 \pm 0.0059$
\\
    $\braket{\wT F_L^{(\ell)}}$ &  
    $0.541  \pm 0.011$  & $0.536  \pm 0.011$ & 
    $0.5491 \pm 0.0102$ & $0.5384 \pm 0.0103$
\\
    $- \braket{S_3^{(\ell)}}$ &
    $0.1350 \pm 0.0036$ & $0.1344 \pm 0.0036$ & 
    $0.1325 \pm 0.0076$ & $0.1452 \pm 0.0079$ 
\\
\hline
\hline
    & \multicolumn{2}{c|}{SM} & \multicolumn{2}{c|}{Fit Belle data}
\\
\hline
    observable & $\Sigma X$  & $\Delta X$
               & $\Sigma X$  & $\Delta X$
\\
\hline\hline
    $10^{14} \cdot \widehat{\Gamma}^{(\ell)}$ [GeV] &
    $(1.467 \pm 0.078)\cdot 10^{3}\cdot|V_{cb}|^2$  & $(-3.80  \pm 0.24)\cdot|V_{cb}|^2$  & 
    $2.1748 \pm 0.0647$ & $-0.0184 \pm 0.0539$ 
\\
    $x_2^{(\ell)}$ &
    $0.0922 \pm 0.0034$ & $(2.272 \pm 0.049)\cdot 10^{-4}$ &
    $0.0936 \pm 0.0017$ & $-0.0040 \pm 0.0029$
\\
    $x_3^{(\ell)}$ &
    $0.1086 \pm 0.0030$ & $(3.119 \pm 0.056)\cdot 10^{-4}$ & 
    $0.1133 \pm 0.0018$ & $-0.0025 \pm 0.0033$
\\
    $x_4^{(\ell)}$ &
    $0.1162 \pm 0.0020$ & $(3.211 \pm 0.087)\cdot 10^{-4}$ & 
    $0.1196 \pm 0.0021$ & $+0.0024 \pm 0.0038$
\\
    $x_5^{(\ell)}$ & 
    $0.1182 \pm 0.0010$ & $(3.10 \pm 0.13)\cdot 10^{-4}$ & 
    $0.1220 \pm 0.0024$ & $+0.0043 \pm 0.0046$
\\
    $x_6^{(\ell)}$ &
    $0.1161 \pm 0.0006$ & $(2.84 \pm 0.17)\cdot 10^{-4}$ & 
    $0.1151 \pm 0.0027$ & $-0.0035 \pm 0.0052$
\\
    $x_7^{(\ell)}$ & 
    $0.1109 \pm 0.0015$ & $(2.47 \pm 0.21)\cdot 10^{-4}$ & 
    $0.1112 \pm 0.0029$ & $+0.0066 \pm 0.0056$
\\
    $x_8^{(\ell)}$ &
    $0.1032 \pm 0.0025$ & $(1.99 \pm 0.24)\cdot 10^{-4}$ & 
    $0.0993 \pm 0.0029$ & $-0.0103 \pm 0.0054$
\\
    $x_9^{(\ell)}$     &
    $0.0935 \pm 0.0033$  & $(1.37 \pm 0.27)\cdot 10^{-4}$  & 
    $0.0923 \pm 0.0026$ & $-0.0000 \pm 0.0052$
\\
    $x_{10}^{(\ell)}$ &
    $0.0819 \pm 0.0040$ & $(2.21 \pm 0.15)\cdot 10^{-4}$  & 
    $0.0768 \pm 0.0025$ & $+0.0019 \pm 0.0044$
\\
    $\braket{F_L^{(\ell)}}$ &
    $0.541  \pm 0.011$  & $(5.43 \pm 0.36)\cdot 10^{-4}$ & 
    $0.5303 \pm 0.0035$ & $-0.0065 \pm 0.0059$
\\
    $\braket{\Afb^{(\ell)}}$ & 
    $0.201  \pm 0.012$  & $(-5.33 \pm 0.24)\cdot 10^{-3}$ & 
    $0.2125 \pm 0.0047$ & $+0.0349 \pm 0.0089$ 
\\
    $\braket{\wT F_L^{(\ell)}}$ &  
    $0.539  \pm 0.011$  & $(-5.20 \pm 0.30)\cdot 10^{-3}$ & 
    $0.5437 \pm 0.0074$ & $-0.0107 \pm 0.0142$
\\
    $- \braket{S_3^{(\ell)}}$ &
    $0.1347 \pm 0.0036$ & $(5.81 \pm 0.22)\cdot 10^{-4}$ & 
    $0.1388 \pm 0.0055$ & $+0.0127 \pm 0.0109$
\\
\hline
\end{tabular}
\renewcommand{\arraystretch}{1.0}
\caption{The SM predictions of observables for $\ell = e$ and $\ell = \mu$,
  using the form-factor results \cite{Bordone:2019guc}, together with
  their values obtained from our fit to the Belle data \cite{Waheed:2018djm}.
  For the prediction of the  total rate, we leave the value of $|V_{cb}|$
  unspecified.
}
\label{tab:SM-predictions}
\end{table}

\begin{figure}
\includegraphics[width=.32\textwidth]{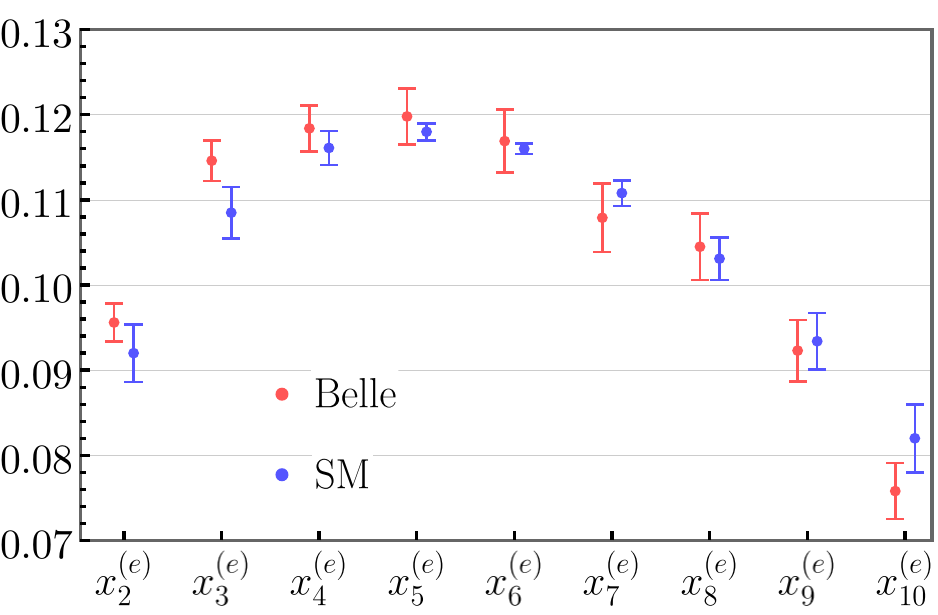}
\includegraphics[width=.32\textwidth]{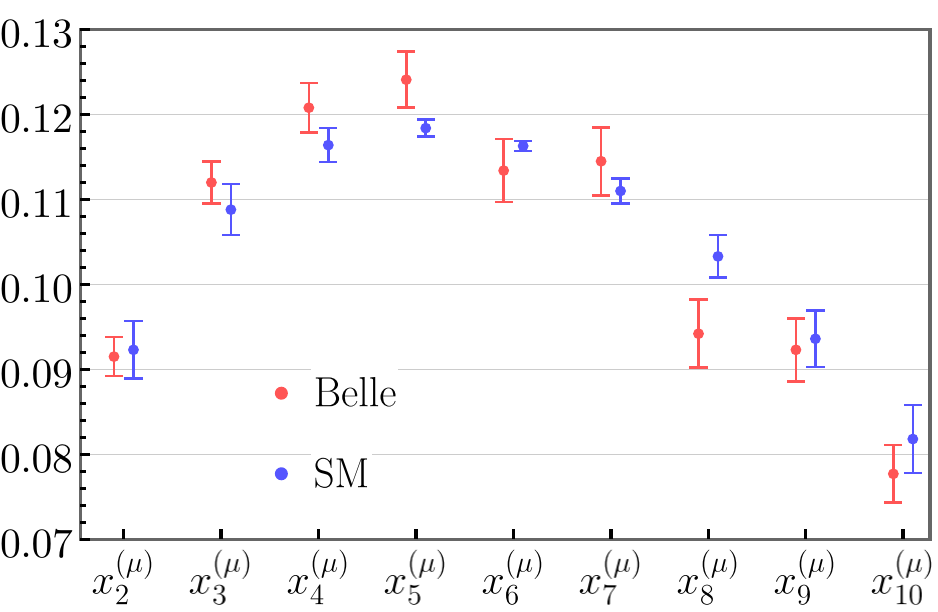}
\includegraphics[width=.32\textwidth]{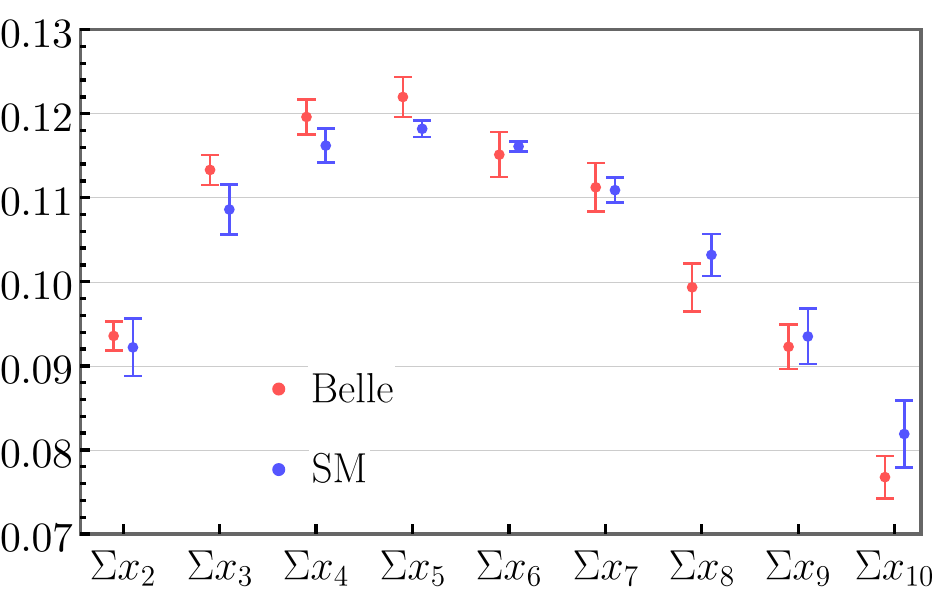}
\\[0.2cm]
\hskip 0.01\textwidth
\includegraphics[width=.33\textwidth]{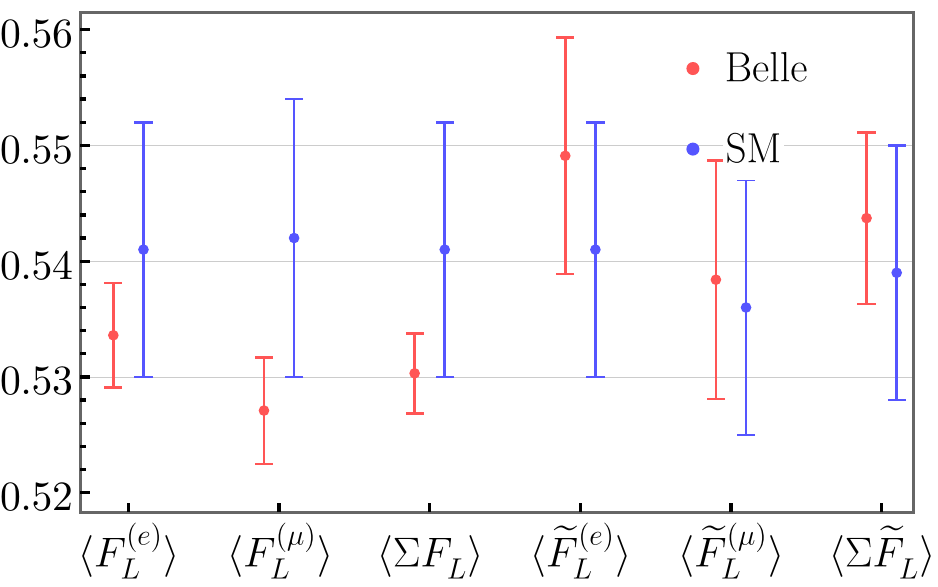}
\hskip 0.18\textwidth
\includegraphics[width=.33\textwidth]{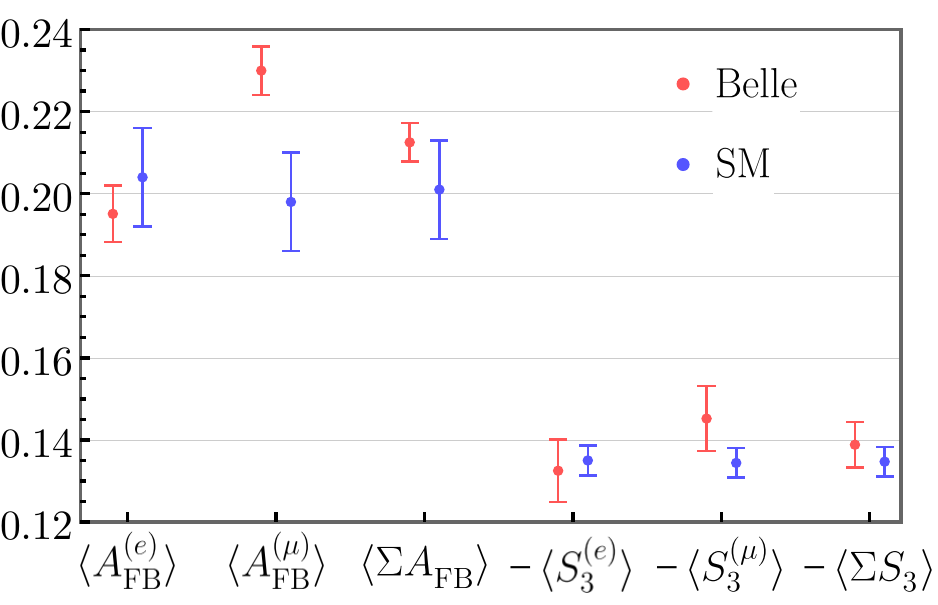}
\\[0.2cm]
\includegraphics[width=.34\textwidth]{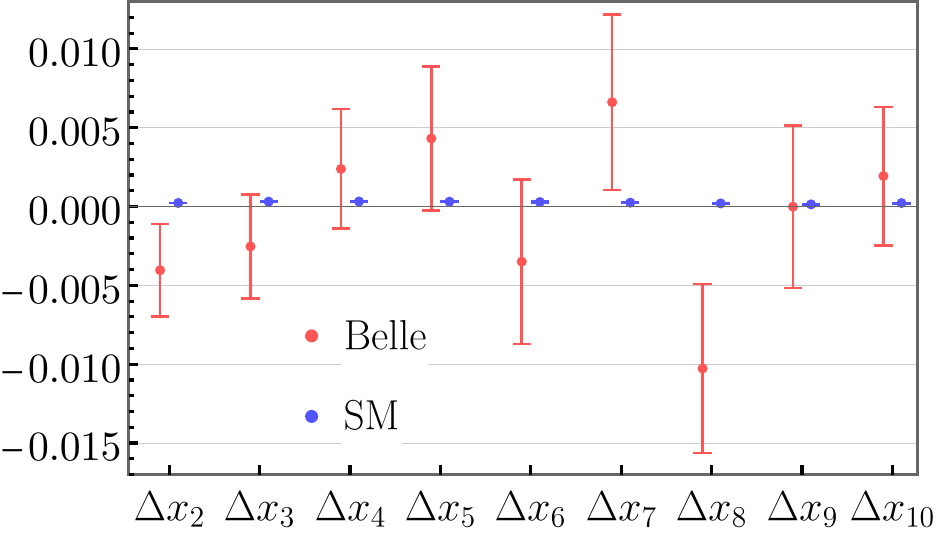}
\hskip 0.18\textwidth
\includegraphics[width=.34\textwidth]{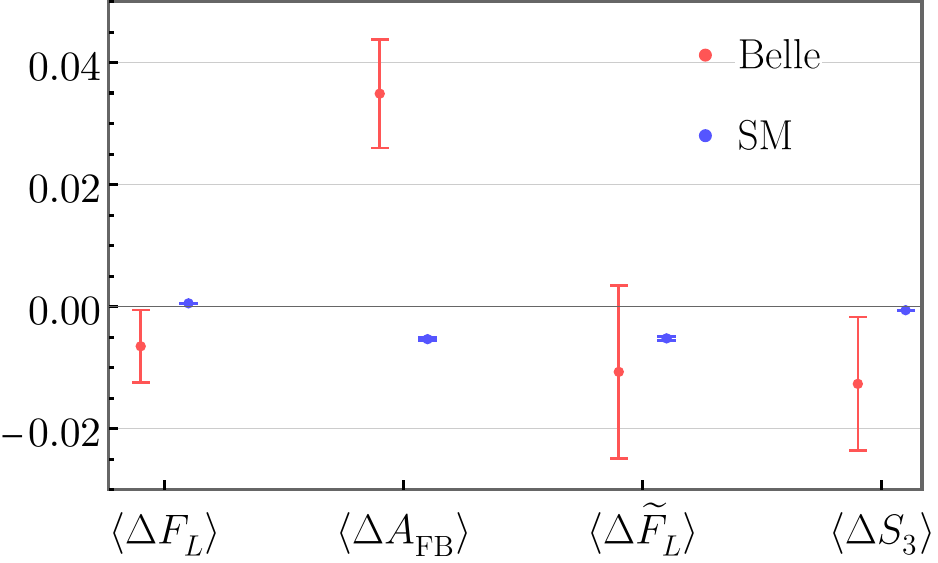}
\caption{\label{fig:x} The measured $x_i^{(\ell)}$ from Belle versus
    the SM predictions for $\ell = e$ [left], $\ell = \mu$ [middle]
    and the lepton-flavour averaged $\Sigma x_i$ [right], and
    also shown the differences $\Delta x_i$ [lower]. The numbers
    are collected in \reftab{SM-predictions}.
}
\end{figure}

We perform two types of fits with our approach to test the stability of
the results:
\begin{enumerate}
\item 
    a simple $\chi^2$ fit,
\item
    a fit using pseudo-Monte Carlo techniques, following the procedure
    described in Ref.~\cite{Ferlewicz:2020lxm},
\end{enumerate}
both using the full $80\times 80$ covariance matrix. In addition, we have
applied a correction to the systematic correlations for d'Agostini bias
\cite{DAgostini:1993arp}, following the procedure described in Ref.~\cite{Jung:2018lfu}.

We find the results of the two fits to be virtually identical.
In Ref.~\cite{Ferlewicz:2020lxm} the authors observe that in their
joint fit of $V_{cb}$ and form-factor parameters the two procedures
produce markedly different results. They conclude that this difference is due
to the large correlations present in the experimental data and that
the usage of the pseudo-Monte Carlo technique is mandatory for phenomenological analyses.
Our findings are in stark contrast to this conclusion and indicate instead
that large correlations alone are not the cause for this difference.
Our
interpretation is that the observed difference is related to the form-factor
parameters entering non-linearly in the fit of Ref.~\cite{Ferlewicz:2020lxm},
while our angular observables and $x_i^{(\ell)}$ parameters enter bilinearly.
It is worth emphasizing in this context that
\begin{itemize}
\item 
    our fit results are extremely well described by Gaussian distributions;
    and that
\item
    the correlations between our fit parameters are much smaller than the
    ones present in the $80\times 80$ matrix describing the bin contents.
\end{itemize}
As a consequence, we do not distinguish between the results from the
two fit procedures in the following.\\[-0.3cm]

The fit results for our parameters as defined in \eqs{eq:obs-set-1}{eq:obs-set-2}
are listed in \reftab{SM-predictions} and shown in \reffig{x}. At the best-fit point we
find $\chi^2 = 48.9$ for $80 -  2\times 14 = 52$ degrees of freedom (dof),
indicating a good fit.\footnote{The dof would be $46$ when considering that
3~bins should be linearly dependent among the 40~bins per lepton flavour,
still indicating a good fit.} This suggests that the assumption
of a pure $P$-wave $D\pi$ final state is well justified.\\[-0.3cm]

In both \reftab{SM-predictions} and \reffig{x} we juxtapose the fit results
with their corresponding SM predictions. The latter depend on the $\bar{B}\to D^*$
form factors. Here, we use the form-factor determinations from Refs.~\cite{Bordone:2019guc, Bordone:2019vic}.
All SM predictions are obtained using the EOS software \cite{EOS}. The EOS code for the computation
of $\bar{B}\to D^*\ell\bar\nu$ observables has been independently checked.
We also predict the ratio $R_{e/\mu}$ in the SM and obtain:
\begin{equation}
    R_{e/\mu} = 1.0026 \pm 0.0001\,,
\end{equation}
which does \emph{not} include possible structure-dependent QED corrections.\\[-0.3cm]

We emphasize that the predictions \cite{Bordone:2019guc, Bordone:2019vic} of the
$\bar{B}\to D^*$ form factors are conservative in that the corresponding uncertainties
include higher-order contributions in the heavy-quark expansion. Furthermore
they rely only on theory input from various sources,
\emph{i.e.} no experimental input has been used for their determination.
Note that $|V_{cb}|$ cancels in the predictions for the normalized bins
$x_i^{(\ell)}$ as well as in the angular observables; only the total
decay rate is proportional to $|V_{cb}|^2$. Moreover, theoretical uncertainties
of the normalization of the leading hadronic $B\to D^*$ form factor cancel
in the normalized observables. However, we do not include structure-dependent
electromagnetic corrections to the angular distribution. Given the expected precision
of the experimental data and the impact of muon-mass effects as discussed in this
work, we expect that including these effects will become mandatory soon.
\\[-0.3cm]

Before comparing to our numerical SM predictions, we test the qualitative
expectation of approximate lepton-flavour-universality, \emph{i.e.}
$\Delta X\equiv 0$, which does not require a specific form-factor parametrization.
We find that most quantities are
well compatible with lepton-flavour universality, with the exception of
$\braket{\Afb^{(\ell)}}$, which shows a deviation from exact universality
at the $3.9\sigma$ level, to be discussed below.
This strong violation is not readily observable in the 80~bins provided by
the Belle collaboration, but becomes obvious in the results of the fit of the 
non-redundant set of angular observables to the underlying angular distributions, see \reffig{x}.
The violation is further hidden by the fact that the lepton-flavour averaged
data are compatible with the SM expectation.\\[-0.3cm]

\begin{figure}
\includegraphics[width=.45\textwidth]{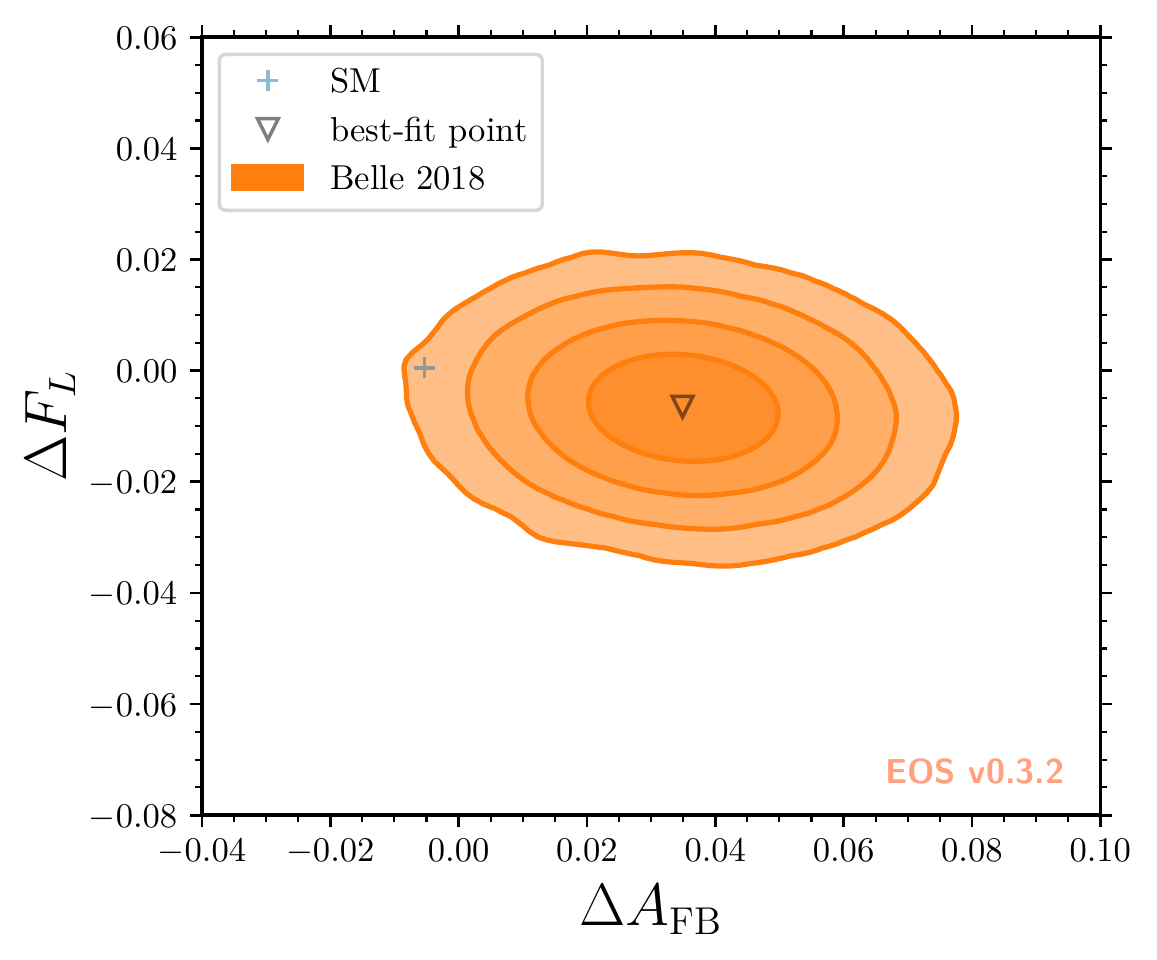}
\includegraphics[width=.45\textwidth]{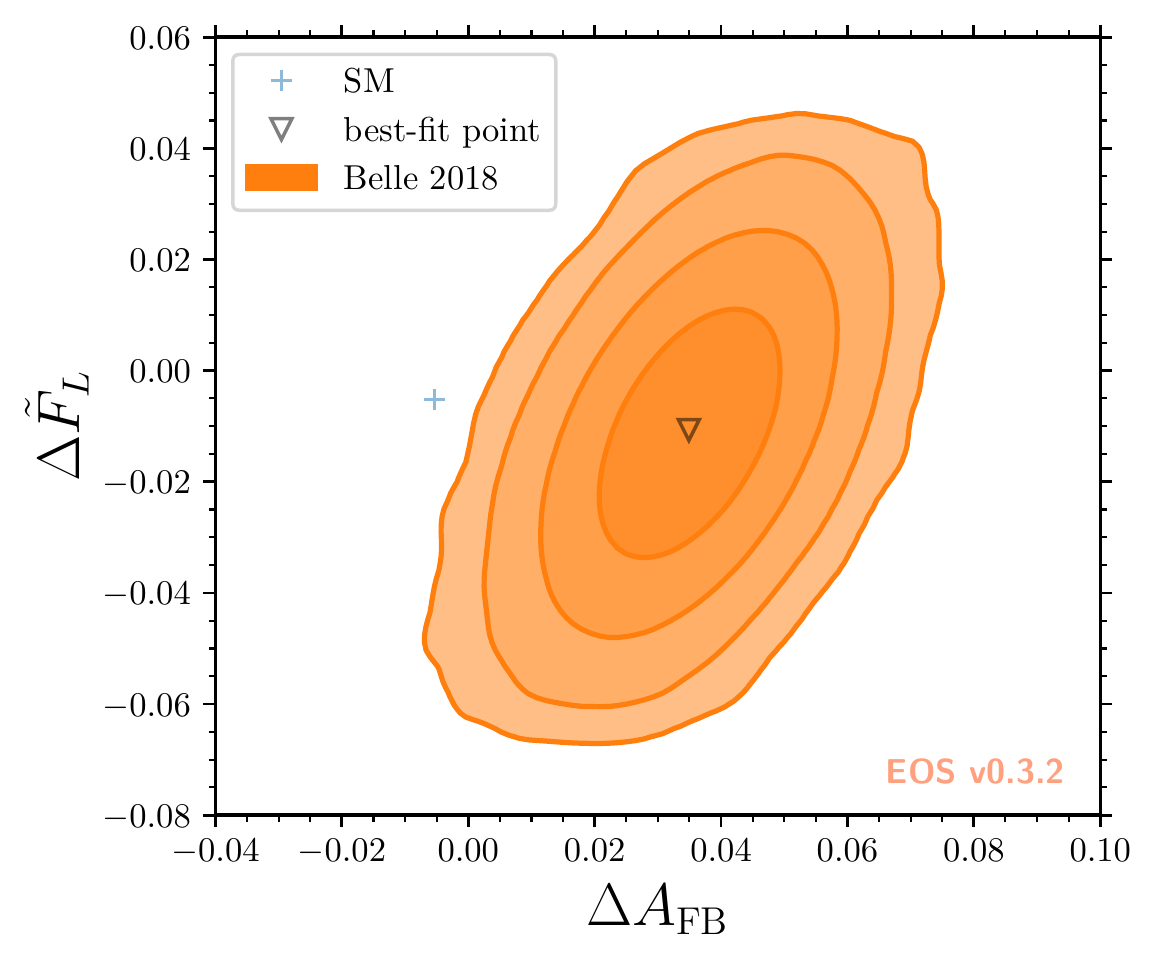}\\
\includegraphics[width=.45\textwidth]{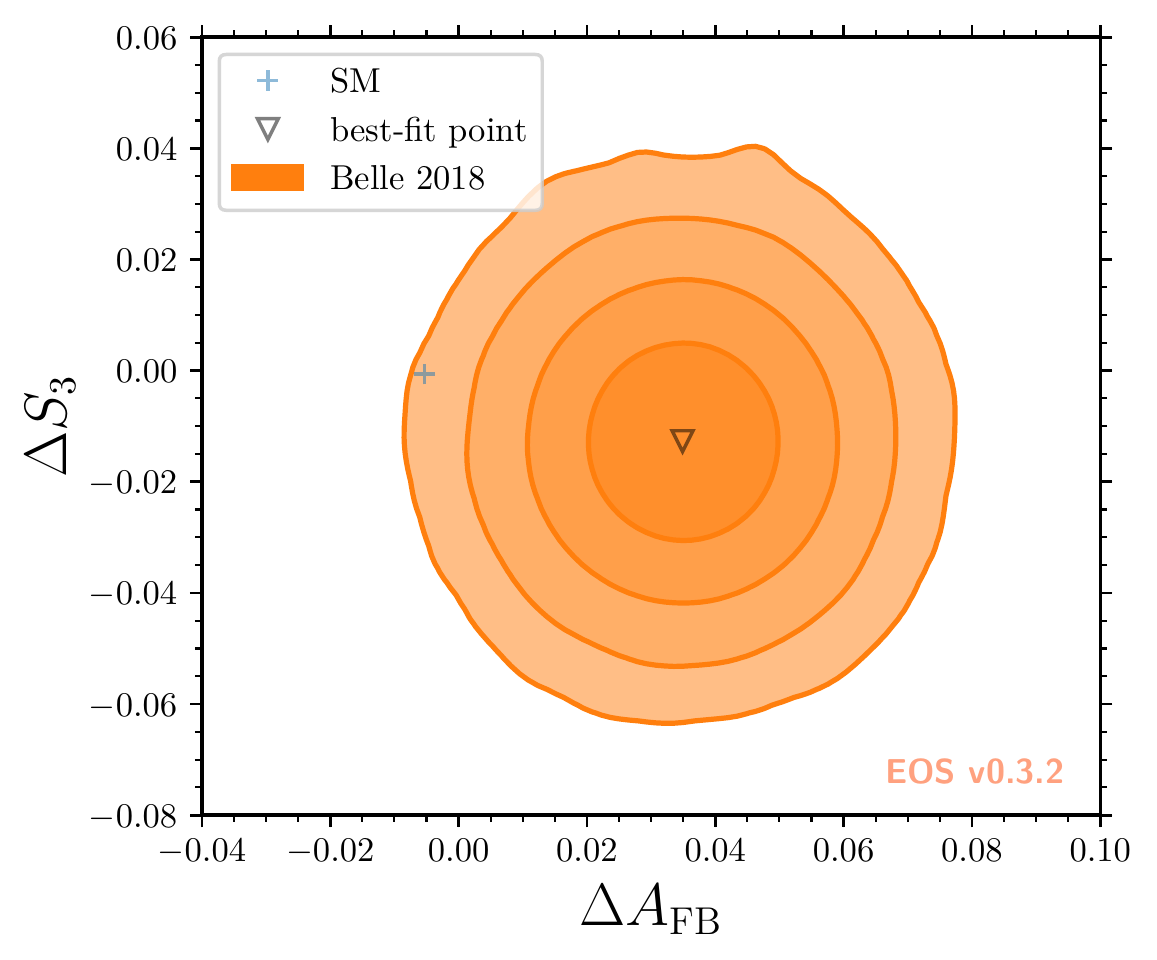}
\includegraphics[width=.45\textwidth]{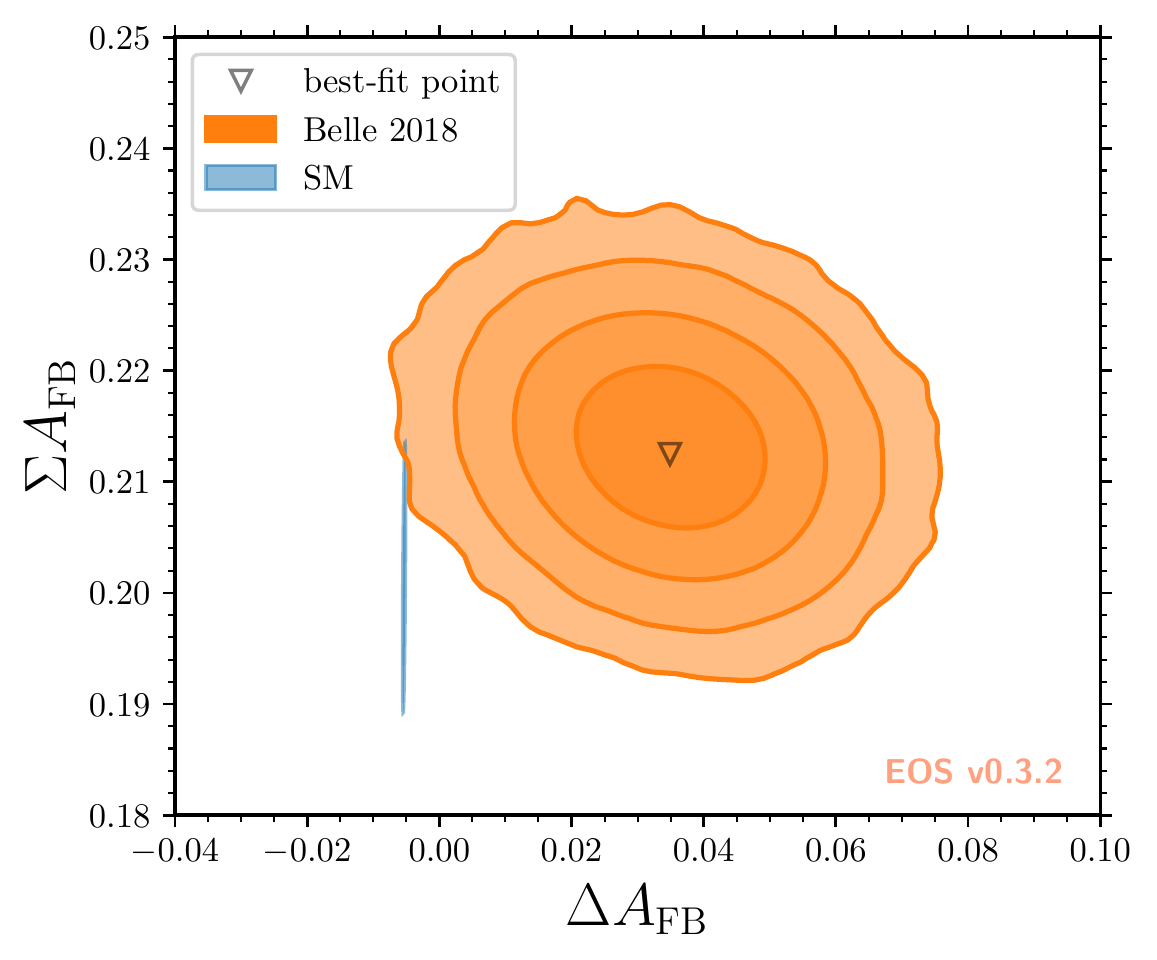}
\caption{\label{fig:DeltaAFBDeltaFLtilde}
    Fit to the Belle data in the planes of $\Delta \Afb = \Afb^{(\mu)} - \Afb^{(e)}$
    vs. $\Delta F_L=F_L^{(\mu)} - F_L^{(e)}$ (top left),
    $\Delta \Afb$ vs. $\Delta\wT F_L=\wT F_L^{(\mu)} - \wT F_L^{(e)}$ (top right),
    $\Delta \Afb$ vs. $\Delta S_3 = S_3^{(\mu)} - S_3^{(e)}$ (bottom left),
    and $\Delta \Afb$ vs. $\Sigma \Afb = (\Afb^{(\mu)} + \Afb^{(e)}) / 2$ (bottom right).
    Contours correspond to $68\%$, $95\%$ $99.7\%$, and $99.99\%$ probability, respectively.
    The ragged outermost contours are artefacts due to lack of samples
    so far in the periphery of the best-fit point.
    The SM predictions based on the form factors
    obtained in Refs.~\cite{Bordone:2019guc,Bordone:2019vic} are shown as blue crosses.
    The SM uncertainties are found to be much smaller than $10^{-2}$ and
    hence negligible, with the exception of the last panel.
    The uncertainty in the $\Delta\Afb$--$\Sigma \Afb$ plane
    is shown as a (highly degenerate) ellipse at the $68\%$ probability level.
}
\end{figure}

In the comparison of our SM predictions with the fit results we find:
\begin{enumerate}
\item 
    As expected, the precision for most normalized quantities is better
    than that for the total rate, typically at the level of a few percent.
    This is true for both the SM predictions and the fit results.
\item
    Overall we find very good agreement of the fit results with our SM
    predictions, as can be seen in \reffig{x}, especially when considering
    the individual lepton species. There are a few smaller differences of roughly
    $1\sigma$, only $\langle\Afb^{(\mu)}\rangle$ shows a tension above
    the $2\sigma$ level.
\item
    The differences  of the lepton-flavour-specific observables, $\Delta X$,
    are predicted
    with very small absolute uncertainties due to the muon-mass suppression.
    Their predictions have similar relative uncertainties as the ones for the
    angular observables themselves. Their absolute values are also very small, with
    $\Delta X/\Sigma X=\mathcal O(\permil)$ in most cases. This can be readily
    understood, since these observables receive only corrections of
    $\mathcal O(m_\mu^2)$ in the SM. The only sizable central values are those
    of $\Delta\Afb$ and $\Delta \wT F_L$, which are slightly enhanced by numerical
    factors. Most importantly, we find that the latter shifts are still small,
    but already comparable to the corresponding experimental uncertainties,
    see \reftab{SM-predictions}. This implies that the muon mass cannot be
    neglected anymore in precision analyses.
\item
    The pattern of the shifts in $\Delta x_i$ is surprising at first sight,
    since $|\Delta x_i|/\Sigma x_i$ is almost constant over the whole range
    of $w$ (or $q^2$), while we argued that the effect scales like
    $(m_\mu/\sqrt{q^2})^2$. This can be understood from the normalization to
    the total rate. The shifts in $\Delta(\Delta \Gamma_i)/\Sigma(\Delta\Gamma_i)$
    scale as expected, from significantly less than $1\permil$ at $w\sim 1$
    (high $q^2$) to $-5\permil$ in the bin with maximal $w$ (lowest $q^2$).
    The shift in the total rate is about $-3\permil$, so normalizing yields
    shifts in $\Delta x_i/\Sigma x_i$ to the range $[-3\permil, 3\permil]$.
\item
    For LFU observables we still find mostly excellent agreement between
    experiment and our SM predictions.
    However, the aforementioned difference between the measurements of
    $\Afb^{(\mu)}$ and $\Afb^{(e)}$ becomes more significant,
    given the smaller absolute
    uncertainty in $\Delta\Afb$ and the fact that the relatively large SM
    prediction carries the opposite sign from the one determined in the fit.
    This quantity differs therefore by approximately $4\sigma$ from its SM
    prediction. In \reffig{DeltaAFBDeltaFLtilde} we show the pair-wise
    2-dimensional best-fit regions of $\Delta \Afb$ with $\Delta F_L$,
    $\Delta \wT F_L$, $\Delta S_3$, and $\Sigma \Afb$. The discrepancy with the
    predictions reaches the $4\,\sigma$ level, compatible with similar levels seen for the 1-dimensional discrepancy for $\Delta \Afb$ in \reftab{SM-predictions}.
\end{enumerate}

These observations mildly depend on the covariance matrix used in the fit.
As stated above, we consider our construction of the $80\times 80$ covariance
matrix reliable to the extent that the data in Ref.~\cite{Waheed:2018djm} are
correct. To make absolutely sure that our assumption regarding the $e-\mu$
correlations is not the reason for the observed discrepancy, we adopt the
following alternative procedure: We determine the $\Afb^{(e)}$ and
$\Afb^{(\mu)}$ with separate statistical and systematic uncertainties
in two separate fits to the lepton-specific data, using the corresponding
$40\times40$ covariance matrices for which we do not have to rely on our
assumption. We then \emph{minimize} the discrepancy with respect to our
(strongly correlated) SM predictions by assuming a diagonal $2\times2$
statistical correlation matrix for $\Afb^{(e)}$ and $\Afb^{(\mu)}$, but
allowing for an arbitrary correlation $\rho \in[-1, 1]$ between the
systematic uncertainties.

We find that the minimal tension with respect to
the SM for the combined $\Afb^{(e)}$, $\Afb^{(\mu)}$ occurs for maximal
anti-correlation ($\rho = -1$), which is not a realistic value.
The correlation determined in the fit to the $80\times80$ covariance matrix is
actually very small.
Adopting nevertheless this most conservative choice of $\rho = -1$ still leads to a tension of $3.6\sigma$. 
We emphasize again that this result is not changed by employing the
pseudo-Monte Carlo approach with Cholesky decomposition for the fit as done
in \cite{Ferlewicz:2020lxm}, nor by the d'Agostini effect (the plots shown in \reffig{x}
include the corresponding shifts). Therefore, even adopting this maximally
conservative procedure, our results amount to evidence for $\mu$-$e$-non-universality
beyond the SM in charged-current $b\to c\ell\nu$ transitions. However, our finding
hinges on the approximate validity of the data and specifically the
correlation matrices given in Ref.~\cite{Waheed:2018djm}.

We also perform a full SM fit to the $2\times 14$ observables in
\reftab{SM-predictions}, including their correlations given in ancillary files attached to the arXiv preprint of this article. Starting from a fit of form-factor parameters from
theory input, only \cite{Bordone:2019guc, Bordone:2019vic}, the inclusion
of the experimental information on these 28 observables increases the
minimal $\chi^2$ by $68.5$, while only $|V_{cb}|$ is introduced as an
additional parameter in the fit. This does indicate a bad fit, with a
$p$ value of $2\times 10^{-5}$, or a tension at the $4.3\sigma$ level. The discrepancy remains
driven by a $\sim 4\sigma$ tension in $\braket{\Afb^{(\mu)}}$ and a $\sim 2\sigma$ tension
in $\braket{\Afb^{(e)}}$. The experimental and theoretical correlations
with other observables play a minor role, see also \reffig{DeltaAFBDeltaFLtilde}.
We note in passing that S-P wave interference cannot affect the numerator of $\Afb$,
and can only decrease the magnitude of $\Afb$ by a coherent contribution to the
denominator \cite{Matias:2012qz}.

We refrain from providing the value of $|V_{cb}|$ from either lepton mode,
which would be compatible with the values obtained from the lepton-flavour
average in Refs.~\cite{Waheed:2018djm, Ferlewicz:2020lxm} and continue to exhibit a substantial
tension with respect to the inclusive determination $|V_{cb}|_{B\to X_c} = (42.00 \pm 0.64)
\cdot 10^{-3}$~\cite{Gambino:2016jkc}.
Given the incompatibility of the data with the SM prediction, we consider it misleading
to use it to extract $|V_{cb}|$.

To summarize, we find in our fits a discrepancy between data and the SM of $\sim 4\sigma$.
This result is stable with respect to the treatment of the d'Agostini bias,
the type of fit we are performing ($\chi^2$ fit vs.~pseudo-Monte Carlo
techniques), and importantly also the precise treatment of the correlations
of the systematic uncertainties between electrons and muons. We reiterate,
however, the concerns discussed in \refsec{BelleData}: the statistical correlation
matrices given in~\cite{Waheed:2018djm} do not seem to be correct, since they
are not singular as they should be, given the performed redistribution of
events to obtain the different single-differential rates. Bearing this caveat
in mind, we still investigate in the following the possibility that the observed
discrepancy is an effect of BSM physics.

%
%
%--------+---------+---------+---------+---------+---------+---------+---------+
\subsection{Possible BSM interpretation}

We consider the possibility that the observed discrepancy is due to BSM physics.
To that aim, we investigate the Lagrangian \refeq{NPlagrangian} in the limit of
lepton-flavour conservation $\ell=\ell'$. From our general analysis in
\refsec{fit-model} we have seen that $\braket{\Afb^{(\ell)}}$ is special in that
it is determined to $\mathcal O(m_\mu)$ only by interference contributions
$\sim \re(C_i^\ell C_j^{\ell*})$, and is the only observable in the
single-differential distributions to which interference terms contribute in the
massless limit. Given the size of the observed effect, $\Delta\Afb /
\Sigma\Afb\sim\mathcal O(10\%)$, a muon-mass suppressed contribution does
not seem likely as its source. This suggests that in order to accommodate
$\Delta\Afb$, the first options to consider are BSM contributions to right-handed
vector operators, to both pseudoscalar and tensor operators, or to left-handed
vector operators. Notably, the first two options correspond to second-order BSM
contributions: for the interference between pseudoscalar and tensor operators
this is obvious. For the right-handed vector operator the interference term
$\re(C_V^\ell C_A^{\ell*}) = |C_{V_R}^\ell|^2-|C_{V_L}^\ell|^2$ is manifestly
second order in the $C_{V_R}^\ell$. For the BSM contributions to the left-handed
vector operator only, the discussion is more involved. The interference terms
$\re(C_V^\ell C_A^{\ell*}) = |C_{V_R}^\ell|^2-|C_{V_L}^\ell|^2$ contain in
principle a linear contribution in $|C_{V_L}^\ell|^2 = |1 + \Delta
C_{V_L}^{\ell,\text{BSM}}|^2$, wherein the $1$ stands for the SM contribution.
However, if $C_{V_L}^\ell$ were the only BSM contribution it would cancel in
all normalized observables. This is not true for the contribution from
right-handed vector operators, the real parts of which, however, enter linearly
in $|C_{A,V}^\ell|^2$. Given the compatibility of all other observables with
the SM, this scenario would therefore require the main contribution to either
have a sizable imaginary part, or specific cancellations with other BSM
contributions, in order not to upset this agreement.\\[-0.3cm]

Taking here the Belle data at face value, we perform fits analogous to the ones
described above, including different sets of BSM contributions. Note that we
keep our description qualitative, since numerical statements are likely to be
upset by an eventual correction of the Belle dataset~\cite{Waheed:2018djm}.
For the same reason we do not perform a combined fit with other
$b\to c\ell\bar\nu$ modes, which would of course be required to confirm the
viability of potential BSM scenarios that resolve the tension in this dataset. \\[-0.3cm]

We find that either contributions from right-handed vector operators, or from
both pseudoscalar and tensor operators are necessary to accommodate the observed
$\Delta \Afb$, confirming our previous considerations. In order to describe the
dataset well with real BSM Wilson coefficients, only, LFUV contributions to both
the right- \emph{and} left-handed vector operators are required.

The three minimal BSM scenarios that fit the present Belle
$\bar B\to D^*\ell\bar\nu$ data~\cite{Waheed:2018djm} can be summarized as follows:
\begin{enumerate}
\item
    $C_{V_R}^\ell\neq 0$: This scenario does require a sizable imaginary part
    (as anticipated above) and LFU violation. The latter fact is interesting,
    since it might point to BSM physics beyond SMEFT \cite{Cata:2015lta}.
    The imaginary part of $C_{V_R}^\ell$ implies that $\braket{A_8^{(\ell)}}$
    and $\braket{A_9^{(\ell)}}$ are sizable. We strongly encourage an
    experimental measurement of these observables. \\[-0.5cm]
\item
    $C_{V_R}^\ell\neq 0$ and $C_{V_L}^\ell\neq 1$: This scenario can obviously
    describe the data well, given that in principle already $C_{V_R}^\ell\neq 0$
    suffices. However, to our surprise it is also compatible with an LFU BSM
    contribution to $C_{V_R}^\ell$, which is required in a SMEFT scenario.
    Enforcing this flavour-universal $C_{V_R}^\ell$, \emph{i.e.,} $C_{V_R}^e = 
    C_{V_R}^\mu$, results in significantly different absolute values and a
    sizable phase difference between $C_{V_L}^e$ and $C_{V_L}^\mu$. Sizable
    $\braket{A_{8,9}^{(\ell)}}$ are also likely in this case, although not strictly
    necessary. It is possible to have all BSM coefficients real, and hence
    $\braket{A_{8,9}^{(\ell)}} = 0$, but only with a  phase between the left-handed coefficients
    $\phi_L=\pi$. This corresponds to a BSM contribution of about twice the
    SM one and is therefore highly fine-tuned.\\[-0.5cm]
\item
    $C_{P}^\ell\neq0$ and $C_{T}^\ell\neq0$: Also this scenario provides a
    good fit to the data, both for complex and real-valued Wilson coefficients.
    The fact that both $C_T^\ell$ and $C_P^\ell$
    are required means that this scenario can be tested by measuring 
    $\braket{S_{6c}^{(\ell)}}$ and $\braket{A_7^{(\ell)}}$, at least one of which is
    expected to show significant differences relative to their SM predictions, which
    are small for $\braket{S_{6c}^{(\ell)}}$ and zero for $\braket{A_7^{(\ell)}}$.
\end{enumerate}
While we do not attempt to include additional datasets as explained above and
therefore cannot quantitatively test specific BSM scenarios, we still observe
a few general features of a possible BSM explanation in the context of the
$B$ anomalies, especially in $b\to c\tau\bar\nu$ transitions:
\begin{enumerate}
\item
    While moderate shifts in one or several Wilson coefficients are required
    to fit the present Belle data \cite{Waheed:2018djm}, the total rates are
    not strongly affected. Hence it is not possible to explain the discrepancy
    in $R(D^{*})$ with these shifts, \emph{i.e.} additional new contributions
    in $b\to c\tau\bar\nu$ coefficients are required to explain
    the deviations of LFU ratios involving $\ell = \tau$ from SM predictions.
\item
    If the observations made here based on the Belle data persist after future
    updates or corrections, they would have strong implications for scenarios
    addressing the $B$ anomalies: Scenarios that only shift $C_{V_L}^\ell$
    would be ruled out, which are currently favoured as simultaneous explanations
    of the $b\to c\tau\bar\nu$ and $b\to s\ell^+\ell^-$ anomalies.
\item
    Based on the picture provided by the observables, one would naively expect
    a hierarchy $\Delta_\mu > \Delta_e$. In light of the more substantial
    deviations in $b\to c\tau\bar\nu$, this could be extended to $\Delta_\tau >
    \Delta_\mu$, which is quite natural in scenarios addressing both $B$ anomalies.
    However, we find that $\Delta_\mu > \Delta_e$ is far from being established
    in our fits at the level of the Wilson coefficients.
\end{enumerate}
There will therefore be far-reaching consequences for the field of particle
physics, should this discrepancy be confirmed.

%
%
%
%--------+---------+---------+---------+---------+---------+---------+---------+
\section{Conclusions}
\label{sec:conclusions}

In this article we pave the way for precision analyses of $b\to c\ell\bar\nu$
processes beyond the assumption of $e-\mu$ universality. This endeavour is
important for the determination of $V_{cb}$ in the Standard Model, a complete
understanding of the weak effective theory (WET) beyond the SM (BSM), and also
to gain new insights into the persistent $b\to c\tau\bar\nu$ anomaly.
We focus on the angular distribution in $\bar{B}\to D^* \ell\bar\nu$ with light
leptons $\ell = e, \mu$ and highlight strategies for improved experimental
analyses.\\[-0.3cm]

We discuss the complete set of CP-even and CP-odd angular observables that arise
from the fully-differential angular distribution of $\bar{B}\to D^* (\to D \pi)\,
\ell\bar\nu$. In particular we discuss the influence of a finite mass of the
charged lepton on these observables in and beyond the SM. We consider in detail
the specific case of four single-differential CP-averaged rates that have been
experimentally analyzed in Refs.~\cite{Waheed:2018djm, Abdesselam:2017kjf}. 
We find that only four flavour-specific angular observables per lepton flavour
are sufficient to describe the three single-differential CP-averaged angular
distributions including arbitrary BSM contributions: the lepton-forward-backward
asymmetry $\Afb^{(\ell)}$, the longitudinal $D^*$-polarization $F_L^{(\ell)}$,
and two further observables $\wT F_L^{(\ell)}$ and $S_3^{(\ell)}$. However, we
find that it is principally not possible to extract the full information on the
BSM contributions to the WET Wilson coefficients for the electron mode when using
only the single-differential CP-averaged rates. For the muon mode, part of that
information enters only muon-mass suppressed, although it can be extracted
without that suppression when considering a different presentation of the data.
We further emphasize the existence non-linear relations between the Wilson
coefficients that allow to test for lepton-flavour violation (LFV) and
right-handed neutrinos.\\[-0.3cm]

The most precise lepton-flavour-specific analysis to date \cite{Waheed:2018djm}
presents the three CP-averaged single-differential angular distributions for
electron and muon flavours separately. Since they depend on only four angular
observables per lepton flavour, the chosen number of kinematic bins is much
larger than necessary. We show that this redundant presentation accidentally
hides tensions between SM predictions and data. We encounter an issue with the
statistical correlation matrices that can only be clarified by the Belle
collaboration. We describe our approach to the combination of statistical and
systematic correlations for the electron and muon datasets and extract the
non-redundant lepton-flavour specific CP-averaged angular observables from the
Belle data. For most of the angular observables we find good agreement with our
up-to-date SM predictions, except for $\Afb^{(\mu)}$. 
The observed tension with the SM predictions is even more pronounced for the
observable $\Delta \Afb \equiv \Afb^{(\mu)} - \Afb^{(e)}$ in which the
correlations of form factors lead to a strong cancellation of uncertainties,
reaching the $4\,\sigma$ level.
We perform numerous checks that this tension is not a result of our specific 
treatment of the data. In particular, even when allowing for arbitrary
systematic correlations between the electron and muon data, we find that this
tension does not drop below $3.6\,\sigma$.
This constitutes evidence for lepton-flavour universality violation.\\[-0.3cm]

We continue by investigating in a qualitative manner the most economic BSM
scenarios that can potentially explain the observed tensions. To this end,
we assume lepton-flavour conservation, but allow for lepton-flavour non-universality
in the WET description. We find that either right-handed vector operators
or both pseudoscalar and tensor operators are necessary to accomodate the
observed tension. If only right-handed vector operators are present, large
imaginary parts in the Wilson coefficients are necessary. As a consequence,
the CP-odd angular observables $A_{8,9}^{(\ell)}$ would be expected to deviate
sizably from their SM predictions. A solution with purely real-valued Wilson
coefficients appears only as a highly fine-tuned solution in a combined scenario
with left- and right-handed vector operators. For the combination of pseudoscalar
and tensor operators, we do not find the necessity of sizable imaginary parts.
In this case, $\braket{S_{6c}^{(\ell)}}$ or $\braket{A_7^{(\ell)}}$
are expected to show significant differences relative to their SM predictions.
None of these three scenarios coincides with the preferred explanation of
the $b\to c \tau\bar\nu$ anomaly.\\[-0.3cm]

Given the far-reaching consequences of our findings, we consider it essential
that the Belle collaboration reviews --- and if need-be corrects --- the
published dataset from Ref.~\cite{Waheed:2018djm}. Without such scrutiny, we
cannot determine the impact of the identified issues on results inferred from
the data. We strongly recommend that
future measurements separate between the two light-lepton flavours in a transparent
way. This is also important for the comparison with existing and upcoming
LHCb analyses, which focus on the muon mode, only.

%
%  Acknowledgments
%
%--------+---------+---------+---------+---------+---------+---------+---------+
\acknowledgments

We are very grateful to David Straub for early discussions of the Belle data and
their possible interpretation in a BSM fit. We also thank Paolo Gambino for
useful discussions. We are grateful to a number of
members of the Belle collaboration for help with the interpretation of the Belle
data, especially Eiasha Waheed, but also Florian Bernlochner, Daniel Ferlewicz,
Daniel Greenwald, Thomas Kuhr, Christoph Schwanda, and Phillip Urquijo.

The work of CB is supported by the Deutsche Forschungsgemeinschaft (DFG, German
Research Foundation) under grant BO-4535/1-1.
The work of MB and MJ is supported by the Italian Ministry of Research (MIUR) under
grant PRIN 20172LNEEZ.
The work of NG is partially supported by DFG under grant 396021762 -- TRR 257
``Particle Physics Phenomenology after the Higgs Discovery''.
The work of DvD is supported by the DFG within the Emmy Noether Programme
under grant DY130/1-1 and by the NSFC and the DFG through the funds provided
to the Sino-German Collaborative
Research Center TRR110 ``Symmetries and the Emergence of Structure in QCD''
(NSFC Grant No. 12070131001, DFG Project-ID 196253076 - TRR 110).
This research was supported by the Cluster of Excellence ``ORIGINS'' and the
Munich Institute for Astro- and Particle Physics (MIAPP) which are funded by
the DFG under Germany's Excellence Strategy -- EXC-2094 -- 390783311.

%--------+---------+---------+---------+---------+---------+---------+---------+
%
%  Appendix
%
%--------+---------+---------+---------+---------+---------+---------+---------+
\appendix

%
%
%
%--------+---------+---------+---------+---------+---------+---------+---------+
\section{Angular Distribution}
\label{app:AngularDistribution}

The complete angular distribution of $\bar B\to D^* (\to D \pi)\, \ell\bar\nu$,
assuming a purely P-wave $D\pi$ state, has been derived in Ref.~\cite{Duraisamy:2014sna},
see also Ref.~\cite{Ivanov:2016qtw}, with previous partial results throughout
the literature \cite{Korner:1989qb,Hagiwara:1989gza,Tanaka:1994ay,Biancofiore:2013ki,Duraisamy:2013pia,Fajfer:2012vx,Tanaka:2012nw}.
In our convention this angular distribution reads:
\begin{equation} 
    \label{eq:ang-dist}
\begin{aligned}
    \frac{8 \pi}{3} 
    \frac{d^4 \Gamma^{(\ell)}}{d q^2\, d\!\cos\thl\, d\!\cos\thD\, d\chi} &
    = \left(J_{1s}^{(\ell)} + J_{2s}^{(\ell)} \cos\!2\thl 
          + J_{6s}^{(\ell)} \cos\thl \right) \sin^2\!\thD
   \\ &
   + \left(J_{1c}^{(\ell)} + J_{2c}^{(\ell)} \cos\!2\thl 
   + J_{6c}^{(\ell)} \cos\thl \right) \cos^2\!\thD  
   \\ &
   + \left(J_3^{(\ell)} \cos 2\chi
   + J_9^{(\ell)} \sin 2\chi \right) \sin^2\!\thD \sin^2\!\thl
   \\ &
   + \left(J_4^{(\ell)} \cos\chi
   + J_8^{(\ell)}  \sin\chi \right) \sin 2\thD \sin 2\thl 
   \\ &
   + \left(J_5^{(\ell)} \cos\chi
   + J_7^{(\ell)} \sin\chi \right) \sin 2\thD \sin\thl \, ,
\end{aligned}
\end{equation}
with twelve angular coefficients $J_i^{(\ell)} = J_i^{(\ell)}(q^2)$, see 
Refs.~\cite{Duraisamy:2014sna, Alguero:2020ukk}. The angles are defined as 
\begin{enumerate}
\item the angle $\thl$ between $\ell$ and the opposite direction of
    flight of the $\bar B$ in the $(\ell\bar\nu)$ center of mass system (cms),
\item the angle $\thD$ between $D$ and the opposite direction of flight
    of the $\bar B$ in the $(D\pi)$ cms, and
\item the angle $\chi$ between the two decay planes spanned by the 3-momenta
    of the $(D\pi)$- and $(\ell\bar\nu)$-systems, in that order.
\end{enumerate}
Throughout we indicate fully $q^2$-integrated quantities $X(q^2)$ with the notation
\begin{align}
  \label{eq:def-braket}
  \braket{X} & \equiv \int dq^2\, X(q^2)\,.
\end{align}
Starting from the $q^2$-integrated decay distribution $d^3\Gamma^{(\ell)} /
d\!\cos\thl\, d\!\cos\thD\, d\chi$ one obtains the integrated decay rate
\begin{align}
\label{eq:Gint}
   \Gamma^{(\ell)} & =
     2 \braket{J_{1s}^{(\ell)}} + \braket{J_{1c}^{(\ell)}} 
   - \frac{1}{3} \big(2 \braket{J_{2s}^{(\ell)}} + \braket{J_{2c}^{(\ell)}} \big) \,,
\end{align}
and the three single-angular differential distributions in
\eqs{eq:dG:dcosthL}{eq:dG:dchi}.
As is customary in the literature, we use in the above the normalized
CP-even and CP-odd angular observables~\cite{Altmannshofer:2008dz}
defined in \refeq{def-S_i-A_i}, where barred quantities refer to
the CP-conjugated decay. While $J_9^{(\ell)}$ can be
non-zero beyond the SM, to leading order in the
weak effective theory $J_9^{(\ell)}$ is purely CP-odd. Hence, non-zero values for
$S_9^{(\ell)} \propto (J_9^{(\ell)} + \bar{J}_9^{(\ell)})$ are highly suppressed.

From the above one finds that the three CP-averaged single-differential
distributions \eqref{eq:dG:dcosthL}--\eqref{eq:dG:dchi} can be described
in terms of only four independent angular observables:
\begin{align}
    \braket{\Afb^{(\ell)}} &
    = \braket{S_{6s}^{(\ell)}} + \frac{\braket{S_{6c}^{(\ell)}}}{2}\,, 
&
    \braket{F_L^{(\ell)}} &
    = \braket{S_{1c}^{(\ell)}} - \frac{\braket{S_{2c}^{(\ell)}}}{3} \,,
&   
    \braket{\wT{F}_L^{(\ell)}} &
    = \frac{1}{3}  - \frac{16}{9} \left(
        \braket{S_{2s}^{(\ell)}} + \frac{\braket{S_{2c}^{(\ell)}}}{2} \right) \,,
\end{align}
and $\braket{S_3^{(\ell)}}$.
We emphasize again that this is true for arbitrary BSM contributions and
independent of the form-factor parametrization. We note furthermore that
in the absence of new tensor or pseudoscalar operators and
in the limit $m_\ell^2 \to 0$ the observable $\wT{F}_L^{(\ell)}$ converges
towards $F_L$, independent of the values of the form factors and the remaining
Wilson coefficients. Hence, the difference between $\wT{F}_L^{(\ell)}$ and
$F_L^{(\ell)}$ is a quasi-nulltest of the SM up to lepton-mass suppressed
effects.

%--------+---------+---------+---------+---------+---------+---------+---------+
%
%  References
%
%--------+---------+---------+---------+---------+---------+---------+---------+

\small

\bibliography{references}

\end{document}